\documentclass[iop]{emulateapj}
\usepackage{natbib}
\usepackage{amssymb,amsmath}
\begin{document}

\title{The Radio Light Curve of the Gamma-Ray Nova in V407~Cyg:\\
Thermal Emission from the Ionized Symbiotic Envelope,\\ Devoured from Within by the Nova Blast}
\author{Laura~Chomiuk\altaffilmark{1,2,3}, Miriam~I.~Krauss\altaffilmark{1}, Michael~P.~Rupen\altaffilmark{1}, Thomas~Nelson\altaffilmark{4}, Nirupam~Roy\altaffilmark{1}, Jennifer~L.~Sokoloski\altaffilmark{5}, Koji~Mukai\altaffilmark{6, 7}, Ulisse~Munari\altaffilmark{8}, Amy~Mioduszewski\altaffilmark{1}, Jennifer~Weston\altaffilmark{5}, Tim~J.~O'Brien\altaffilmark{9}, Stewart~P.~S.~Eyres\altaffilmark{10}, and Michael~F.~Bode\altaffilmark{11}}
\email{chomiuk@pa.msu.edu}
\altaffiltext{1}{National Radio Astronomy Observatory, P.O. Box O, Socorro, NM 87801 USA}
\altaffiltext{2}{Harvard-Smithsonian Center for Astrophysics, 60 Garden Street, Cambridge, MA 02138, USA}
\altaffiltext{3}{Department of Physics and Astronomy, Michigan State University, East Lansing, MI 48824}
\altaffiltext{4}{School of Physics and Astronomy, University of Minnesota, 116 Church St. SE, Minneapolis, MN 55455}
\altaffiltext{5}{Columbia Astrophysics Laboratory, Columbia Univesity, New York, NY 10027, USA}
\altaffiltext{6}{CRESST and X-ray Astrophysics Laboratory, NASA/GSFC, Greenbelt, MD 20771, USA}
\altaffiltext{7}{Center for Space Science and Technology, University of Maryland Baltimore County, Baltimore, MD 21250 USA}
\altaffiltext{8}{INAF Astronomical Observatory of Padova, 36012 Asiago \{VI\}, Italy}
\altaffiltext{9}{Jodrell Bank Centre for Astrophysics, University of Manchester, Manchester M13 9PL, UK}
\altaffiltext{10}{Jeremiah Horrocks Institute, University of Central Lancashire, Preston, PR1 2HE, UK}
\altaffiltext{11}{Astrophysics Research Institute, Liverpool John Moores University, Twelve Quays House, Egerton Wharf, Birkenhead, CH41 1LD, UK}

\begin{abstract}

We present multi-frequency radio observations of the 2010 nova event in the symbiotic binary V407~Cygni, obtained with the Karl G.~Jansky Very Large Array and spanning 1--45 GHz and 17--770 days following discovery. This nova---the first ever detected in gamma rays---shows a radio light curve dominated by the wind of the Mira giant companion, rather than the nova ejecta themselves. The radio luminosity grew as the wind became increasingly ionized by the nova outburst, and faded as the wind was violently heated from within by the nova shock. This study marks the first time that this physical mechanism has been shown to dominate the radio light curve of an astrophysical transient. We do not observe a thermal signature from the nova ejecta or synchrotron emission from the shock, due to the fact that these components were hidden behind the absorbing screen of the Mira wind. 

We estimate a mass loss rate for the Mira wind of $\dot{M}_w \approx 10^{-6}\ {\rm M}_{\odot}\ {\rm yr}^{-1}$. We also present the only radio detection of V407~Cyg before the 2010 nova, gleaned from unpublished 1993 archival VLA data, which shows that the radio luminosity of the Mira wind varies by a factor of $\gtrsim$20 even in quiescence. Although V407~Cyg likely hosts a massive accreting white dwarf, making it a candidate progenitor system for a Type Ia supernova, the dense and radially continuous circumbinary material surrounding V407~Cyg is inconsistent with observational constraints on the environments of most Type Ia supernovae.

\end{abstract}

\begin{figure*}
%\centering
%\begin{center}
\hspace{0.8in}
 \includegraphics[width=12cm, angle=90]{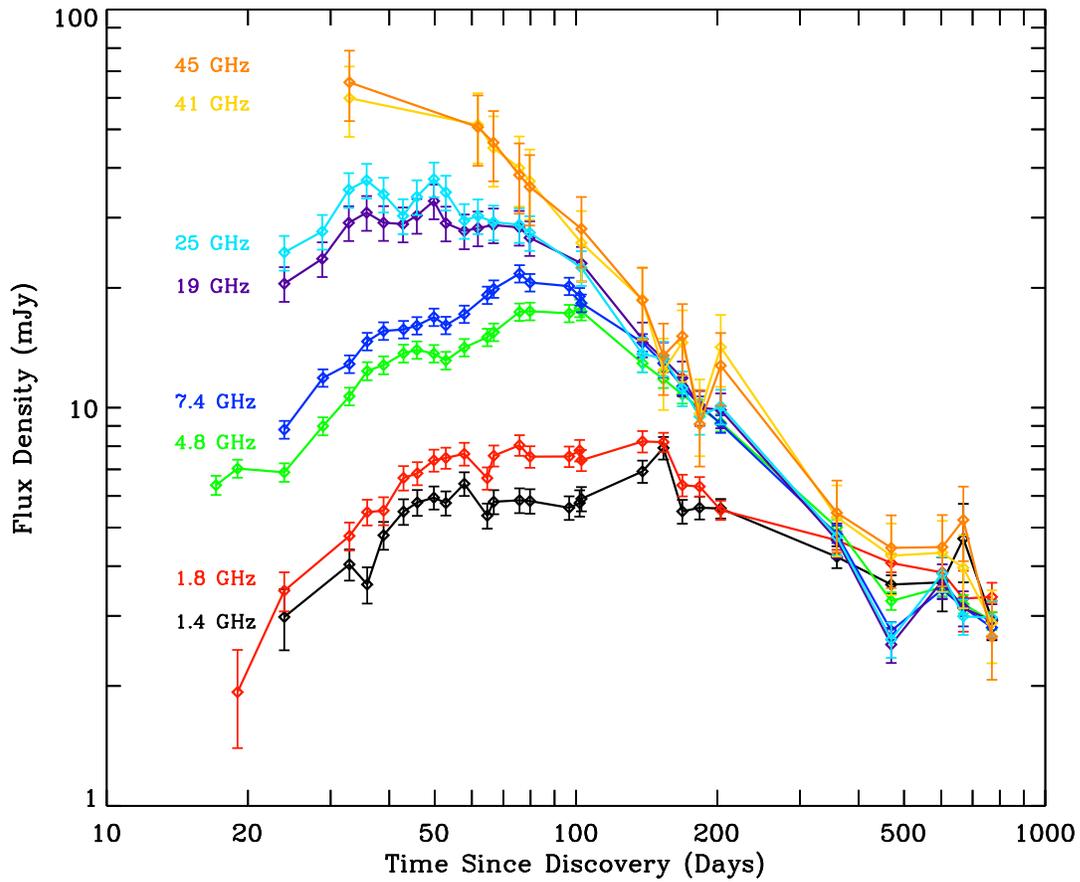}
\caption{Multi-frequency radio light curve for V407~Cyg. A variety of colors denote different frequencies spanning 1.4--45 GHz, as marked on the left side of the plot. The time of optical discovery of the nova event is 2010 March 10.8. }
\label{lc}
%\end{center}
\end{figure*}

\section{Introduction}

On 2010 March 10, the symbiotic star V407~Cygni hosted the first nova explosion ever detected in gamma rays \citep{Cheung_etal10, Abdo_etal10}. This explosion also constitutes the second example of an ``embedded" nova thoroughly observed at a range of wavelengths, presenting an opportunity to trace the interaction of a blastwave with dense circumstellar material in detail (the first example was RS Ophiuchi, with recent outbursts in 1985 and 2006; \citealt{Bode87, Evans_etal08}). Many parallels can be drawn between RS Oph and V407~Cyg \citep{Mikolajewska10, Nelson_etal12}, and together they form a class of intriguing candidates for Type Ia supernova (SN) progenitors \citep{Munari_Renzini92, Hachisu03, Patat_etal11}.

V407~Cyg is a D-type (dust-rich) symbiotic binary featuring a white dwarf (WD) and a Mira giant star in a wide orbit \citep{Munari_etal90}. The giant pulsates with a period of 763 days \citep{Meinunger66, Kolotilov_etal98, Kolotilov_etal03}, the longest known period for a Mira symbiotic system. There is some indication of a 43-year orbital period in the long-term optical light curve \citep{Munari_etal90}, corresponding to a binary separation of 17 AU for a WD mass of 1 M$_{\odot}$ and a giant mass of $1.5$ M$_{\odot}$.

V407~Cyg has undergone periods of increased activity throughout the past several decades, observed as peaks in the optical light curve and brightening emission lines, particularly in the years 1936 and 1998 \citep{Hoffmeister49, Welin73, Munari_etal90, Kolotilov_etal98, Kolotilov_etal03, Tatarnikova_etal03a, Esipov_etal12}. These outbursts likely mark an increase in the accretion rate onto the WD \citep{Kolotilov_etal03}, and so do not qualify as \emph{bona fide} novae, which are powered by a thermonuclear runaway.

However, the 2010 outburst \emph{was} a true nova explosion, displaying an unprecedented brightening of the optical light curve (reaching a B-band magnitude of $\sim$8 mag, as opposed to 13.1 mag in the 1998 outburst; \citealt{Kolotilov_etal03, Nishiyama_etal10, Munari_etal11}) and an expansion velocity of the ejecta of 3,200 km s$^{-1}$ \citep{Buil10, Abdo_etal10}. The unique gamma-ray signature from this explosion has been attributed to particles accelerated by the interaction of the nova shock with the dense Mira wind \citep{Abdo_etal10, Lu_etal11,Hernanz_Tatischeff12, Sitarek_Bednarek12, Martin_Dubus12}. The optical properties of V407~Cyg's nova event have been described by \citet{Munari_etal11} and \citet{Shore_etal11, Shore_etal12}, while the X-ray evolution was studied by \citet{Shore_etal11}, \citet{Orlando_Drake12}, \citet{Schwarz_etal11}, and \citet{Nelson_etal12}. 

This paper presents complementary radio continuum observations of the 2010 outburst, obtained between 1--45 GHz with the Karl G.~Jansky Very Large Array (VLA; \citealt{Perley_etal11}). Before the 2010 outburst, no radio detections of V407~Cyg had ever been reported (although, see Section \ref{hist}), but shortly after the nova explosion V407~Cyg was strongly detected across the centimeter and millimeter bands \citep{Nestoras_etal10, Gawronski_etal10, Pooley10, Bower_etal10, Krauss_etal10}. In Section 2, we present our data and describe the radio light curve. In Section 3, we present a new distance estimate to V407~Cyg, based on optical \ion{Na}{1} D absorption lines. Section 4 considers and dismisses the possibility that the radio emission from V407~Cyg is dominated by expanding thermal ejecta expelled in the nova. In Section 5, we hypothesize that the radio emission is produced by the Mira wind, similar to the radio-emitting ionized envelopes seen around other symbiotic systems \citep{Seaquist_etal84}, but ionized by the nova outburst. In Section 6, we predict the synchrotron signal from V407~Cyg and show that it was significant but hidden behind the absorbing screen of the Mira wind. Finally, the first radio detection of V407~Cyg in quiescence is presented in Section 7, using archival VLA data from 1993. We conclude in Section 8.

\section{The Data}

\subsection{Observational Method}\label{obs}
At the time of the outburst of V407~Cyg, a new correlator was being commissioned as part of the VLA upgrade. We therefore began monitoring V407~Cyg on 2010 April 3.7, approximately one month after discovery (2010 March 10), as some of the newly upgraded facility's first observations \citep{Krauss_etal10}. During each epoch, we attempted to observed at L (1--2 GHz), C (4--8 GHz), K (19--26 GHz), and Q (40--50 GHz) bands, providing coverage across the entire VLA frequency range. Under program AS1039, we obtained data roughly twice a week throughout April and May 2010. The spacing of our epochs gradually widened, and monitoring was continued until April 2012 via programs 10B-233, 11A-254, and 11B-170. Our monitoring of V407~Cyg has now bridged more than an entire VLA configuration cycle, beginning in its most compact D configuration. During the summer (June--August) of 2011, the VLA was in the extended A configuration, and V407~Cyg was spatially resolved. The details of this imaging will be presented in Mioduszewski et al. (2013, in preparation), but flux densities were difficult to derive during this time at some frequencies. 

In a typical epoch, we spent two hours total to observe all four bands, providing $\sim$10--15 minutes on source at each band. At K and Q bands, we obtained improved pointing solutions on a nearby calibrator, and used J2102+4702 for gain calibration. At C band, we used J2048+4310 for gain calibration, and at L band, J2038+5119 and J2052+3635 were used. The absolute flux density scale and bandpasses were calibrated using 3C48. The earlier data (programs AS1039 and 10B-233) were obtained with 256 MHz of bandwidth per band, split between two independently tuneable basebands. Later observations (programs 11A-254 and 11B-170) provided 2 GHz of bandwidth, split between two 1-GHz-wide basebands. 

We also utilized some early data publicly available in the archive under program AL733 (PI Lang), obtained on 2010 March 27.9 and 2010 March 29.8 at L and C bands. This program observed with one-hour-long blocks at each receiver band, and used 256 MHz of bandwidth (at L band, this bandwidth was split between 1.4 and 1.9 GHz, while at C band the basebands were set adjacent to one another at 5 GHz).

Data reduction was carried out using standard routines in CASA and AIPS. Gain curves and weather tables (using the mean of the seasonal average and the day's measurements) were applied. Basic editing and calibration were carried out on each epoch and receiver band separately. Typically, one iteration of phase-only self calibration was applied, and the two basebands were imaged separately. We measured flux densities of V407~Cyg by fitting gaussians to the unresolved point-like source, except during the A-configuration (epoch 26), when two Gaussians were fit and summed together to represent the higher-frequency flux densities. Calibration errors of 5\% were assumed at L and C bands, 10\% at K band, and 20\% at Q band (Q band was particularly plagued by poor receiver performance, a faint gain calibrator, and inclement weather). Measurements are presented in Table \ref{tab:phot}.

Additional radio continuum data were published by \citet{Nestoras_etal10}, \citet{Gawronski_etal10}, \citet{Pooley10}, and \citet{Bower_etal10} in Astronomer's Telegrams. Unfortunately, the facilities and frequencies used for these measurements are heterogeneous, so they are difficult to integrate into our VLA light curve and are not used in our analysis (although a spectral index from the Nestoras et al.~data is listed in Table \ref{tab:spind}).

\begin{figure*}
%\centering
\hspace{0.6in}
\includegraphics[width=13cm, angle=90]{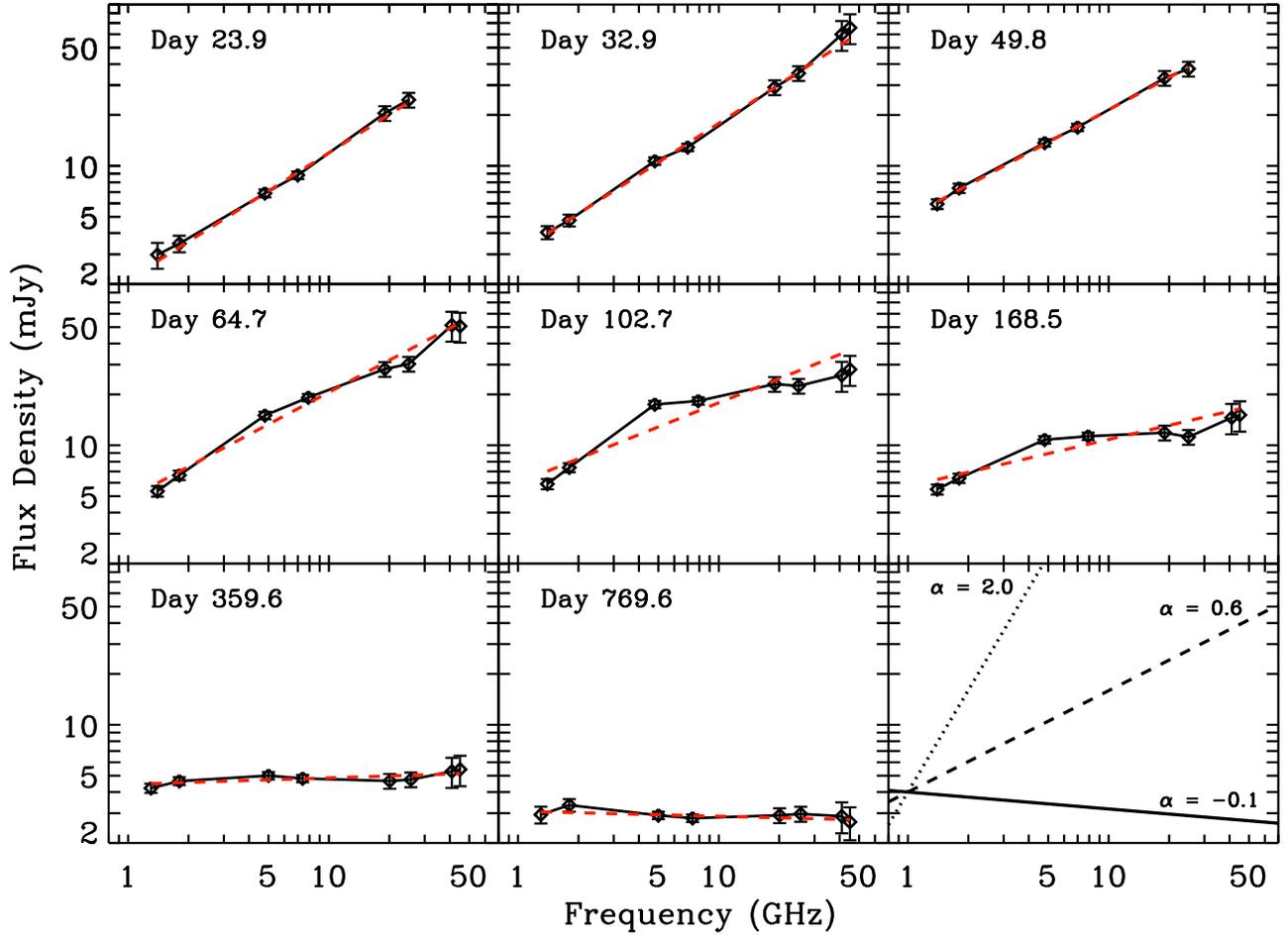}
\caption{Selected radio spectra showing the evolution of V407~Cyg. The 1--45 GHz power law fits described in Table \ref{tab:spind} are marked as dashed red lines. In the bottom-right panel, comparison spectra are shown for three physically significant spectral indices.}
\label{spfit}
\end{figure*}

\subsection{Results}\label{radev}

Our multi-frequency radio light curve for V407~Cyg, plotted in Figure \ref{lc}, shows a gradual rise and decline of radio luminosity over the two years following the nova outburst. Due to observational complications, there were occasionally small shifts in the central frequencies of our bands, as shown in Table \ref{tab:phot}; therefore, for example, the ``7.4 GHz'' light curve actually includes data centered at frequencies between 7.0--7.9 GHz. To minimize dependence on model assumptions, no spectral index correction was applied to our light curve plots to account for this small difference.

The first VLA observations were obtained 17 days after outburst at only 5 GHz, and V407~Cyg is already radio bright at this time ($\sim$7 mJy). Overall, the light curve rises and declines, with the higher frequencies peaking at larger luminosities and declining first. At 45 GHz, V407~Cyg is already fading by the time of our first reliable Q-band observation on Day 33. At frequencies between 4.9 and 25 GHz, the source continues to brighten for a month or two longer, before turning over and declining. At 1.4 and 1.8 GHz, the flux density rises until Day $\sim$70, and then remains roughly constant at this peak flux density for another $\sim$100 days before gently declining. By Day 300, the flux densities at all frequencies converge to $\sim$3.5 mJy and then remain roughly constant for the next year, through the time of writing of this paper.

The entire radio evolution of V407~Cyg is relatively gentle and slow, showing gradual rises and broad peaks in the 1.4--25 GHz light curves. Typical expectations for novae attribute the radio emission to expanding thermal ejecta, with luminosity increasing in proportion to area while it is optically thick (\citealt{Seaquist_Bode08}; Section \ref{ejecta})---so, $L_{\nu} \propto t^2$. However, the rise of the radio light curves in V407~Cyg is significantly slower than this, scaling as $L_{\nu} \propto t^{0.5-1.0}$.

Even at early times, the spectral index $\alpha$ (defined as $S_{\nu} \propto \nu^{\alpha}$) of V407~Cyg is significantly shallower than the $\alpha = 2$ predicted for optically-thick thermal emission \citep[e.g.,][]{Seaquist_Bode08}. The report of \citet{Nestoras_etal10} and the early VLA data from program AL733 indicate that the spectrum may be steepest at early times, with $\alpha \gtrsim$ 1.0 (see Table \ref{tab:spind}). By the time our full VLA monitoring effort began on Day 24, the spectral index was measured to be $\alpha = 0.7-0.8$. 

On Day 24, the spectrum could be fit with a single power law of $\alpha = 0.75$ between 1.4 and 49 GHz (Figure \ref{spfit}). This spectral characterization remained roughly constant through Day $\sim$70, with some evidence for gentle flattening with time to $\alpha \approx 0.6$. After Day 70, the 25 GHz light curve clearly turns over and the spectra show flattening above $\sim$20 GHz. By Day 100, the spectral flattening has cascaded down to $\sim$6 GHz, and by Day 200 the entire spectrum is consistent with a flatter power law of $\alpha = 0.25$. At late times ($>$ 350 days after outburst), the spectral index is even flatter, consistent with an optically-thin thermal spectrum ($\alpha = -0.1$; Figure \ref{spfit}).

\begin{figure}
\centering
\includegraphics[width=8.5cm]{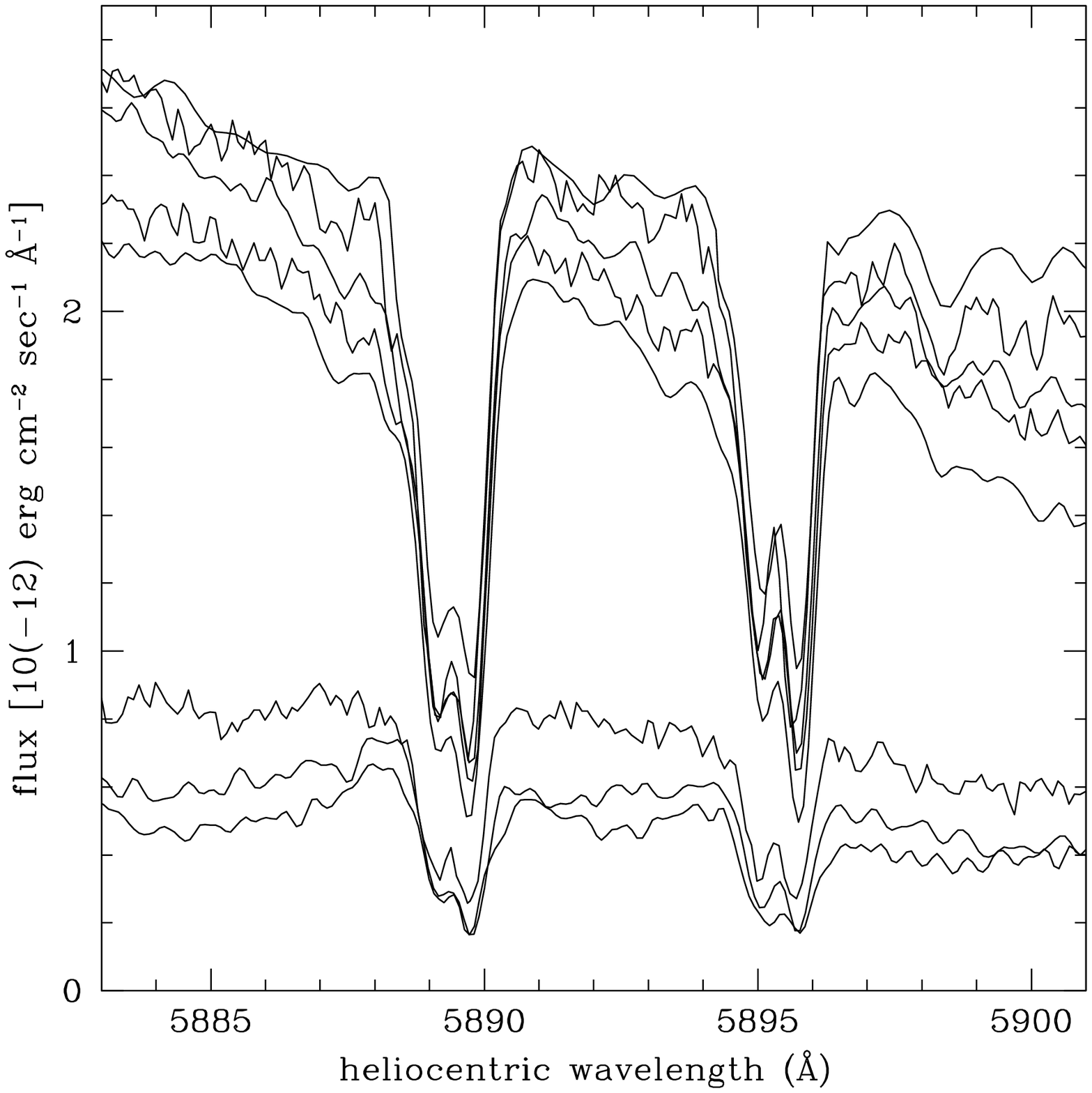}
\caption{Detail of echelle spectra around the \ion{Na}{1} D absorption lines. We plot eight distinct spectra, observed at a range of times after the nova outburst, with the absorption lines superimposed on the time-variable red wing of the \ion{He}{1} $\lambda$5976 line. The \ion{Na}{1} D absorption in all spectra shows a constant two-component structure.}
\label{na}
\end{figure}

\section{Distance to V407~Cyg}
Our interpretation of the radio light curve depends on the distance to V407~Cyg, which is poorly known. Estimates in the literature range from 1.7 kpc \citep{Kolotilov_etal03} to 2.7 kpc \citep{Munari_etal90}, all using the period-luminosity relation for Miras. However, the period-luminosity relation is not constrained for the very long period of V407~Cyg \citep[e.g.,][]{Whitelock_etal08}, and the relation must be widely extrapolated to derive a distance. In addition, \citet{Tatarnikova_etal03b} identified lithium in the atmosphere of V407~Cyg. Li-rich AGB stars have been noted to lie above the period-luminosity relation, displaying anomalously high luminosities for a given period \citep{Whitelock_etal03}. Therefore, distances to V407~Cyg derived from the period-luminosity relationship are highly uncertain and likely underestimate the true distance.
%Here, we use the \ion{Na}{1} D absorption profiles to show that the distance to V407~Cyg is likely $\gtrsim$3.0 kpc.

To provide additional constraints on the distance using interstellar \ion{Na}{1} D absorption lines, we revisited the echelle spectra published by \citet{Munari_etal11}. Two velocity components were observed that were constant in time as the 2010 outburst evolved (Figure \ref{na}), indicating an interstellar foreground origin. They have heliocentric velocities of $-45.7\pm1.1$ km s$^{-1}$ and $-10.9\pm1.0$ km s$^{-1}$. In the local standard of rest (LSR), the corresponding velocities are $-29.9$ km s$^{-1}$ and $+4.9$ km s$^{-1}$ (for Galactic coordinates $l = 86.98, b = -0.48$).

\begin{figure*}
\centering
%\hspace{1.5in}
\includegraphics[width=14cm, angle=90]{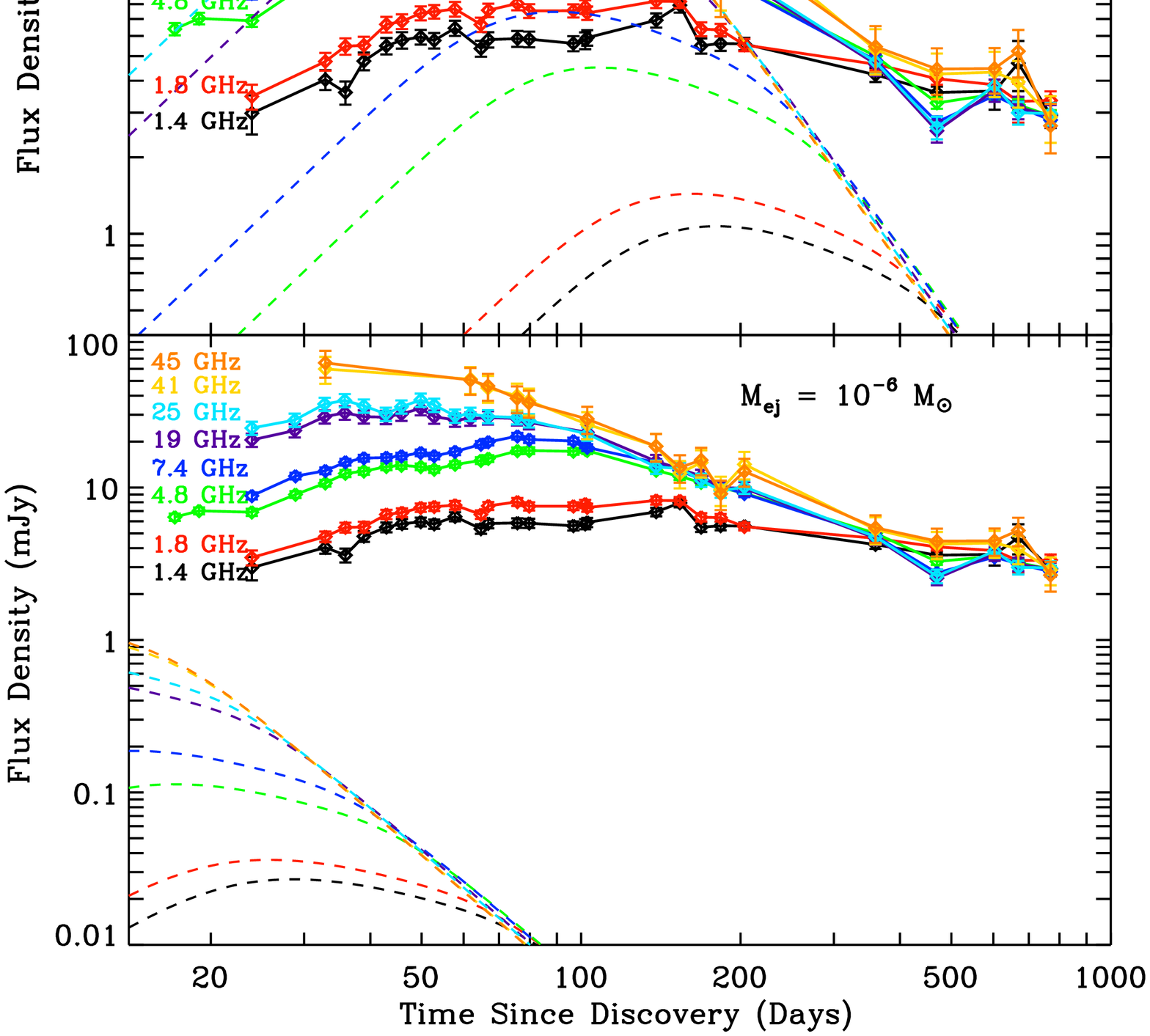}
\caption{Models for the multi-frequency radio emission produced by thermal expanding nova ejecta are superimposed as dotted lines on our VLA light curves for V407~Cyg. The models assume a ``Hubble flow" distribution of ejecta, a maximum velocity of 3,200 km s$^{-1}$, a distance of 3.0 kpc, and a ratio of $v_{\rm max}/v_{\rm min}$ = 0.1. The top panel shows a model for a massive ejection that fits the data relatively well (M$_{\rm ej} = 1 \times 10^{-4}$ M$_{\odot}$, T$_{\rm ej} = 4 \times 10^{4}$ K), while the bottom panel shows a model consistent with multi-wavelength observations of V407~Cyg and expectations for recurrent novae (M$_{\rm ej} = 1 \times 10^{-6}$ M$_{\odot}$, T$_{\rm ej} = 1 \times 10^{4}$ K). Note that the vertical scales of the top and bottom plots differ.}
\label{lcmod}
\end{figure*}

We use the Galactic rotation curve of \citet{Brand_Blitz93} to translate the LSR velocities of the absorbing components to distances of 3.0 kpc and 1.5 kpc, respectively. Both components are strong and nearly saturated (Figure \ref{na}), indicating a significant amount of absorbing material. If the Milky Way's neutral ISM is approximated as a slab extending 150 pc from the midplane, our line of sight will run through it for $\sim$3.5 kpc, consistent with the existence of absorption from a distance of 3.0 kpc. This line of sight crosses the Orion Spur spiral arm at a distance of $\sim$1.5 kpc and meets the large Perseus spiral arm at a distance of ~3 kpc \citep{Churchwell_etal09}. Therefore, the interstellar \ion{Na}{1} D absorption implies a distance $\gtrsim$3.0 kpc for V407~Cyg.

Several other authors have hypothesized that the more blue-shifted absorption component ($v_{\rm LSR} = -30$ km s$^{-1}$) is due to the circumstellar envelope of V407~Cyg itself \citep{Tatarnikova_etal03a, Shore_etal11}. If this hypothesis were accurate, we would only be able to constrain the distance of V407~Cyg to $\gtrsim$1.5 kpc. Tatarnikova et al.~point out that the more blue-shifted \ion{Na}{1} D component has a very similar radial velocity as the \ion{Ca}{1} and \ion{Li}{1} absorption features, which presumably arise in the Mira's atmosphere. However, we observe that the more blue-shifted absorption component did not vary in strength relative to the less blue-shifted component over the course of the nova outburst, implying that both absorption components originate in the interstellar---not the circumstellar---medium. In addition, while the narrow absorption component of H$\alpha$ varies in its central wavelength throughout the nova outburst (it shifts blueward with time), the \ion{Na}{1} absorption is remarkably constant in its central wavelength (Munari et al.~2012, in preparation). We therefore conclude that the correspondence in velocity between the $v_{\rm LSR} = -30$ km s$^{-1}$ \ion{Na}{1} D absorption component and stellar features is a coincidence, and both \ion{Na}{1} D absorption components are interstellar.

Hence, we find that V407~Cyg is located $\gtrsim$3.0 kpc away. We will take a distance at the lower end of this range, $D = 3.0$ kpc, throughout this paper, as it is consistent with most other works on the 2010 nova, which assume a distance of 2.7 kpc. However, we maintain that the distance to V407~Cyg is still uncertain. Where appropriate throughout our work, we note how a change in the distance would affect our results.

\section{Thermal Nova Ejecta as a Potential Source of Radio Emission from V407~Cyg} \label{ejecta}

Here, we compare the multi-frequency radio light curve of V407~Cyg with a model that has historically explained the radio emission from most classical novae: expanding, thermal ejecta spherically expelled by the nova. We show that if this mechanism is the dominant source of radio luminosity, the nova ejection in V407~Cyg must have been massive ($10^{-5}-10^{-4}$ M$_{\odot}$)---in conflict with significant observational evidence that the ejecta were low in mass ($10^{-7}-10^{-6}$ M$_{\odot}$).

In the past, most measurements of radio emission from novae were consistent with thermal bremsstrahlung from a $\sim$10$^{4}$ K ionized sphere that is expanding in size, dropping in density, and gradually transitioning from optically thick to optically thin \citep[e.g.,][]{Seaquist_Palimaka77, Hjellming_etal79}. The radio emission has been interpreted as the expanding nova ejecta, and from basic assumptions about its density profile and from simple modeling of its expansion, the ejecta mass and energy can be estimated \citep{Seaquist_Bode08}. This scenario has mostly been applied to classical novae with minimal circumbinary material surrounding them, and indeed, the radio emission from the only other well-studied nova in a symbiotic system, RS Oph, has proven much more complex \citep{Hjellming_etal86, O'Brien_etal06, Kantharia_etal07, Rupen_etal08, Sokoloski_etal08, Eyres_etal09}. Still, here we begin with the simplest explanation for radio emission in V407~Cyg, and compare our VLA light curve with expectations for the expanding nova ejecta alone.

To first order, most radio light curves of classical novae can be fit with a simple model wherein the ejecta are instantaneously ejected and homologously expanding. The model requires a maximum and minimum velocity ($v_{\rm max}$ and $v_{\rm min}$) as input for this ``Hubble flow", and the ejecta are assumed to be distributed in a thick expanding shell with a $\rho \propto r^{-2}$ radial density profile. In this simple model, the ejecta are also assumed to be isothermal, with an electron temperature $\sim 10^{4}$ K. Additional complications can certainly be added to this model, like clumping, temperature increases, and non-spherical geometry \citep[e.g.,][]{Hjellming96, Heywood04, Roy_etal12}. However, in most cases, the simplest model has been sufficient to approximate observed light curves.

In Figure \ref{lcmod}, we superimpose such models on our VLA light curve for V407~Cyg to test if the ejecta might account for the majority of the radio emission. We assume a distance of 3.0 kpc, $v_{\rm max} = 3,200$ km s$^{-1}$ \citep{Munari_etal11, Shore_etal11}, and a thick shell with $v_{\rm min}/v_{\rm max} = 0.1$. Although no Hubble-flow model provides a satisfactory fit to the data at all frequencies, the top panel shows a relatively good fit to the higher frequencies and earlier times. This model requires a slightly higher electron temperature than typically assumed, $T_{\rm ej} \approx 4 \times 10^{4}$ K (although consistent with observations of some novae; e.g., \citealt{Eyres_etal96}), to fit the high flux densities and relatively early peak at 41/45 GHz. It also requires a massive ejection (M$_{\rm ej} \approx 1 \times 10^{-4}$ M$_{\odot}$) to achieve the high measured radio flux densities. 

Such a large ejecta mass is typical of classical novae, but contradicts optical and X-ray observations which imply that the nova in V407~Cyg was a relatively low-mass ejection, more typical of recurrent novae than classical novae. The optical light curve declined by three magnitudes in the V-band over 24 days \citep{Munari_etal11}. According to nova models, such a rapid decline should only occur if $M_{\rm ej} < 10^{-6}$ M$_{\odot}$ \citep{Yaron_etal05}. In addition, \citet{Mikolajewska10} points out that V407~Cyg's light curve matches the light curve of RS~Oph's 2006 nova event remarkably well; estimates of the ejecta mass for RS Oph range across $\sim 10^{-7}-10^{-6}$ M$_{\odot}$ \citep{O'Brien_etal92, Sokoloski_etal06, Drake_etal09, Eyres_etal09, Vaytet_etal11}. Additional support for a low ejecta mass is lent by a soft X-ray component, observed in V407~Cyg between Days $\sim$15--50 and identified as thermal emission from residual nuclear burning on the WD surface \citep{Schwarz_etal11,Nelson_etal12}. The speed with which the ejecta become optically thin is also confirmed by optical spectroscopy, where high-ionization lines (e.g., [\ion{Fe}{11}]) first appear around Day 7 and subsequently brighten \citep{Munari_etal10}. The rapid ``turn-on" time of soft X-ray emission and coronal lines imply a low ejecta mass of $\sim 10^{-6}-10^{-7}$ M$_{\odot}$, assuming an ejecta velocity of 3,200 km s$^{-1}$ (\citealt{Schwarz_etal11}).

A small ejecta mass of M$_{\rm ej} = 1 \times 10^{-6}$ M$_{\odot}$ yields the model shown in the bottom panel of Figure \ref{lcmod}. This model underpredicts the radio flux densities by two orders of magnitude. The model light curves also evolve extremely quickly, with the predicted 5 GHz curve peaking $\sim 20$ days after the start of outburst rather than at Day $\sim 100$ as observed.

Even if the ejection in V407~Cyg was massive, serious problems remain in fitting the radio spectral energy distribution with the model pictured in the top panel of Figure \ref{lcmod},. The figure illustrates this point; the model matches the 45 GHz light curve rather well up until Day $\sim$160, but offsets between data and models grow toward lower frequencies. At 1.8 GHz, there is roughly an order of magnitude discrepancy between the observed and model light curves. This discrepancy is caused by the model prediction of an optically-thick thermal spectrum at early times ($\alpha$ = 2), which is much steeper than the $\alpha = 0.6-1.0$ that we observe (Table \ref{tab:spind}). The model also predicts that, once the thermal emission becomes optically thin, it will rapidly fade. In contrast, we have observed a late-time plateau in the light curve of V407~Cyg with a duration of $>$300 days (continuing through present) during the time when the spectral index implies optically-thin emission. 

The simple thermal model assumes that the ejecta are not clumpy, but smooth in density with a volume filling factor of $f = 1$. However, observations of novae reveal that the ejecta can be quite clumpy, with $f = 10^{-5}-10^{-1}$ \citep[e.g.,][]{Saizar_Ferland94, Mason_etal05, Ederoclite_etal06, Shara_etal12}. In a simple model, for isothermal ejecta with a single filling factor describing all times and radii, clumpiness simply increases the radio luminosity by a constant factor over the uniform-density case, with no changes to the time evolution or radio spectrum \citep{Abbott_etal81, Nugis_etal98, Heywood04}. This basic case is likely an over-simplification, and more complex modelling of the radio emission from multi-temperature and time-variable clumped ejecta is needed. However, it is difficult to imagine that the addition of such complications could slow the evolution of the low-mass model light curve by the factor of $\sim$4 needed to match the data. Therefore, while clumping might serve to increase the radio luminosity from our M$_{\rm ej} = 1 \times 10^{-6}$ M$_{\odot}$ model to bring it into better alignment with observations, clumping can not account for the dramatic discrepancies in light curve timescales and spectral indices.

Of course, the nova ejecta in V407~Cyg are decelerating with time as they interact with the wind from the Mira giant companion (clearly seen in optical line profiles; \citealt{Munari_etal11, Shore_etal11}). In the process of decelerating, the outer ejecta will be swept up into a relatively thin and dense shell; the radio emission will therefore remain optically thick longer than in the constant expansion model, and then transition more quickly to an optically thin state. Such an evolution would only exacerbate the disagreement we already see between the data and models of expanding nova ejecta----namely, an observed spectral index at early times which is much flatter than $\alpha = 2$ and a more-gradual-than expected rise and decline in the observed light curve. This consideration serves as a reminder that significant circumbinary material surrounds the nova, which may itself contribute to the radio emission (see the next Section).

The required ejecta mass shrinks, but remains quite high ($\sim$4$\times 10^{-5}$ M$_{\odot}$) if the distance is small, $\sim$1.5 kpc. Independent of distance and clumping, a model with M$_{\rm ej} \lesssim 10^{-6}$ M$_{\odot}$ peaks far too quickly and declines too rapidly, compared with our observed light curve for V407~Cyg. In view of the poor fits to expanding thermal ejecta and the well-established presence of circumbinary material around V407~Cyg, we now proceed to consider other physical explanations for the radio emission from V407~Cyg's 2010 nova explosion.

\begin{figure*}
\centering
\includegraphics[width=15cm, angle=90]{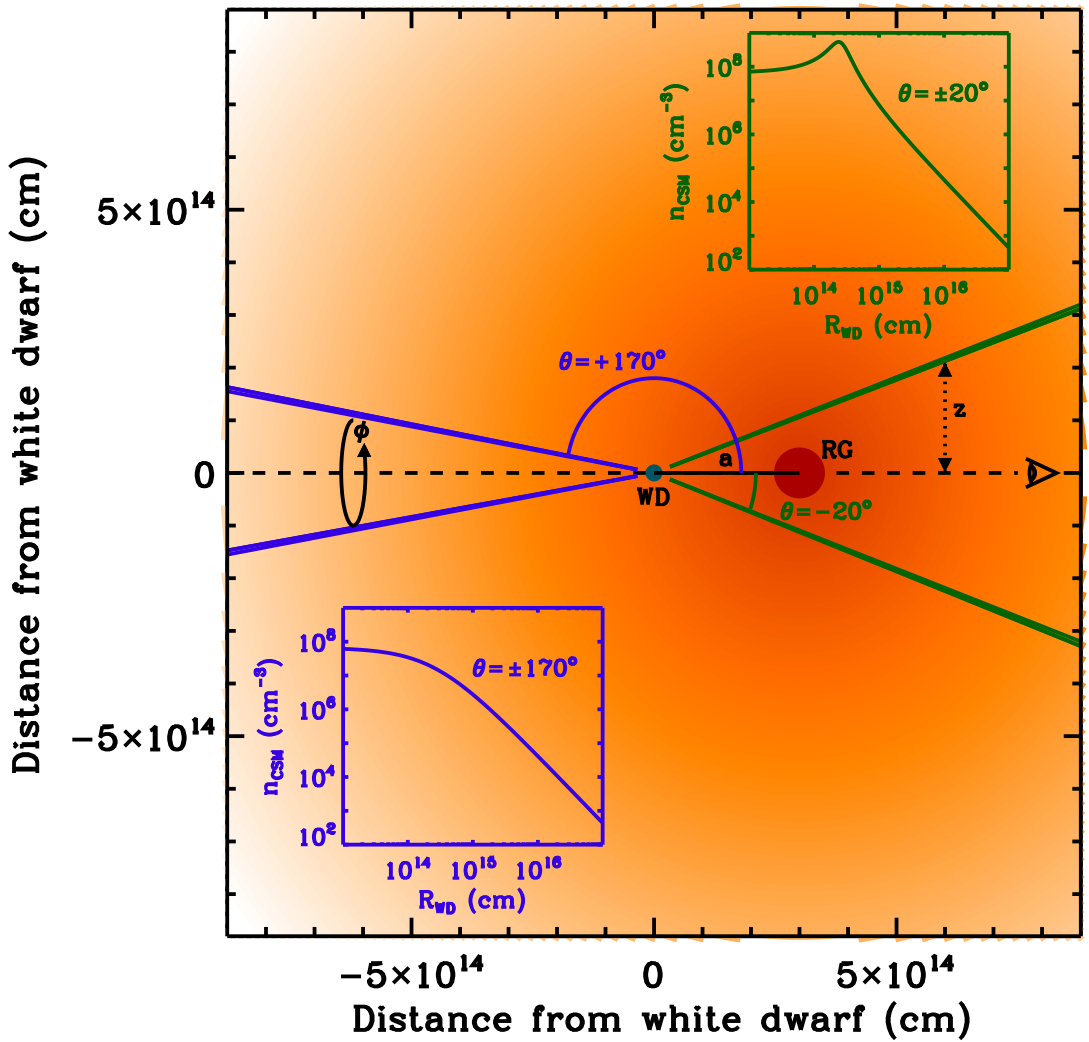}
\caption{The set-up for our model of V407~Cyg. The size of the blue and red circles, denoting the WD and Mira giant respectively, are larger than to-scale for clarity. A wind emanates from the Mira giant, depicted as an orange radial gradient. Along a given polar angle $\theta$, the radial density profile of the circumstellar material is approximated as independent of azimuthal angle $\phi$. The radial density profiles depend on the separation between the WD and Mira giant ($a$) and the wind mass-loss parameters $\dot{M}_{w}$ and $v_w$. Two such density profiles are plotted, for $\theta = \pm20^{\circ}$ and $\theta = \pm170^{\circ}$, assuming $\dot{M}_{w} = 2 \times 10^{-6}$ M$_{\odot}$ yr$^{-1}$, $v_w = 10$ km s$^{-1}$, and $a = 17$ AU. We calculate $\tau = 1$ surfaces and determine their radii from the WD at each of an array of impact parameters ($z$), assuming that the system is being viewed in the orbital plane, with the Mira giant between the viewer and the WD (from the right, in this figure).}
\label{ill}
\end{figure*}

\section{Radio Emission From An Ionized Mira Wind, Shocked From Within}\label{wind}

We now consider the contribution of the Mira wind to the radio emission from V407~Cyg. The ionized envelopes surrounding symbiotic stars emit thermal radio emission; their luminosities scale with the wind density and the degree of ionization, but span the range of radio luminosities observed from V407~Cyg after the 2010 nova \citep{Seaquist_Taylor90}. We show that the Mira wind in V407~Cyg can produce the bulk of the radio luminosity, as it becomes ionized by the hot WD. The radio luminosity grows while the wind is becoming more ionized, and declines as this ionized wind gradually recombines and becomes swept up by the nova shock.

\subsection{A Quantitative Model}

Here, we expand on the simple model introduced in \citet{Nelson_etal12}, which was originally intended to describe the evolution of the nova blastwave in the asymmetric circumbinary medium surrounding the WD in V407~Cyg. We solve for the ionization of the wind as a function of radius, polar angle from the WD, and time. We compare the location of the ionization front with the location of the nova shock front (also a function of polar angle from the WD and time), to estimate the portion of the wind that has been shock-heated and no longer emits in the radio band. Then, assuming a viewing direction, we can calculate the radio flux density and model the radio light curve. A similar mechanism has been suggested by \citet{Kenny_Taylor05} to explain the radio flux densities and images of symbiotic stars in outburst; they model these systems as an ionized, fast, hot wind emanating from the WD, colliding with a slower, cooler, partially-neutral wind from the red giant.

V407~Cyg is an inherently asymmetric system; photons emanating from the WD in the direction of the giant encounter significantly denser material that those traveling away from the red giant. Our model accounts for this asymmetry by splitting the environment of the binary into slices of polar angle ($\theta$) centered on the WD position (Figure \ref{ill}). Assuming spherical symmetry around the Mira giant, a given slice can be described with a single radial profile of circumstellar density, independent of azimuthal angle ($\phi$) and time. The density profile in a given polar-angle slice depends only on the binary separation $a$, the mass-loss rate of the wind $\dot{M}_w$, and the wind velocity $v_{w}$. Relative to the Mira giant, the density of the spherically symmetric circumbinary material is:
\begin{equation}\label{eq:dens}
n_{\rm CSM}(r_{\rm RG}) = {{\dot{M}_w} \over {4 \pi \mu m_{H} v_w}} r_{\rm RG}^{-2}
\end{equation}
where $n_{CSM}$ is the number density of particles in the wind at a radius $r_{\rm RG}$ from the Mira giant; $\mu$ is the mean molecular weight and $m_{H}$ is the hydrogen mass. Throughout this paper, we quote $\dot{M}_w$ assuming $v_w$ = 10 km s$^{-1}$, as estimated for V407~Cyg from optical P~Cygni line profiles \citep{Abdo_etal10}. 
Relative to the WD, the circumbinary medium is asymmetric as a function of $\theta$:
\begin{equation}
\begin{split}
n_{\rm CSM}(r_{\rm WD}) = & {{\dot{M}_w} \over {4 \pi \mu m_{H} v_w}} \times \\
&  \left[(a - r_{\rm WD}\,{\rm cos} \theta)^2 + (r_{\rm WD}\,{\rm sin} \theta)^2\right]^{-1}
\end{split}
\end{equation}
where $\theta = 0$ is the direction pointing from the WD towards the Mira giant and $r_{\rm WD}$ is the distance from the WD. Radial density profiles for two different values of $\theta$ (one toward the Mira giant and the other away) are plotted in Figure \ref{ill}. Regions of the Mira wind away from the giant will become ionized more quickly than toward the giant (illustrated by comparison of our model's density profiles for the $\theta = \pm170^{\circ}$ and $\theta = \pm20^{\circ}$ directions, shown in Figure \ref{ill}). Also, the nova blastwave will experience less deceleration in the $\theta = \pm170^{\circ}$ direction than in the $\theta = \pm20^{\circ}$ direction.

\subsubsection{Photoionization Model}
We assume that, at the time of nova outburst, the Mira wind is completely neutral. We then input a time-variable flux of ionizing photons ($N_{u}(t)$, in units of photons s$^{-1}$), assumed to be emanating from the WD as a result of nuclear burning on its surface. Our selections for this important input are described more below, in Section \ref{ionass}. The fraction of the ionizing flux that enters a given polar-angle slice centered at $\theta$ is:
\begin{equation}\label{fvol}
f_{\rm vol} = {{1}\over{2} }{\rm cos} \left(\theta - {{\Delta \theta}\over{2}}\right) - {{1}\over{2}} {\rm cos} \left(\theta + {{\Delta \theta}\over{2}}\right).
\end{equation}
The width of the polar-angle slice is $\Delta \theta$, and set to 1 deg in our model (for a total of 180 slices).

At a given time step, a given polar angle slice is split into radial bins of width $\Delta r$; starting at small $r_{\rm WD}$ and traveling outward, we calculate how much ionizing flux reaches a given radial bin and the ionization fraction $x(r,\theta,t)$ in that bin,. The ionizing flux at the WD photosphere is $S_{u}(r=0,\theta,t) = f_{\rm vol}\, N_{u}(t)$. The ionizing flux reaching radial bin $(r,\theta,t)$ is $S_{u}(r-1,\theta,t)$, while the ionizing flux escaping bin $(r,\theta,t)$, after ionizing the neutral atoms in this bin, is $S_{u}(r,\theta,t)$. The relationship between incident and emerging radiation is:
\begin{equation}\label{ibal}
\begin{split}
&S_{u}(r,\theta,t)  =  S_{u}(r-1,\theta,t) \\
& - \left[{{4 \pi \left[1-x(r,\theta,t-1)\right] n_{\rm csm}(r,\theta)\, r_{\rm WD}^2\, \Delta r\, f_{\rm vol}(\theta)} \over {\Delta t}}\right] \\
& - \left[4 \pi\, x(r,\theta,t-1)^2\, n_{\rm csm}(r,\theta)^2\, \alpha_{(2)}\, r_{\rm WD}^2\, \Delta r\, f_{\rm vol}(\theta)\right]
\end{split}
\end{equation}
All parameters are in cgs units. The second term on the right hand side represents the number of neutral atoms in this radial bin available for absorbing the ionizing flux. The third term is the rate of recombinations in this bin---newly neutral atoms that will also absorb $S_{u}(r-1,\theta,t)$. Here, $\alpha_{(2)}$ is the recombination coefficient of hydrogen, excluding recombinations to the $n=1$ level. At a temperature of $10^{4}$ K, $\alpha_{(2)} = 2.76 \times 10^{-13}$ cm$^{3}$ s$^{-1}$ \citep{Spitzer78}. This value is conservative, as recombination plays a relatively weak role in the evolution of radio emission from V407~Cyg; hotter assumed temperatures would give lower values of $\alpha_{(2)}$ and even weaker effects from recombination.

We assume that the ionization front is sharp; all atoms in a bin must become ionized ($x(r,\theta,t) = 1$) before the residual ionizing flux is allowed to escape to the next radial bin. In support of this assumption, we only include photons with energies between the Lyman limit and 5 times the Lyman limit (13.6--66.5 eV) in the count of ionizing photons. The hardest photons are excluded due to their small cross sections for ionizing hydrogen; they presumably serve to ionize metals, although we include no treatment of metals in our simple model. If the ionizing flux reaching a bin is not sufficient to ionize all of the atoms in this bin (in other words, if $S_{u}(r,\theta,t)$ in Equation \ref{ibal} is negative), the fractional ionization of this bin is calculated as:
\begin{equation}
\begin{split}
& x(r,\theta,t) = x(r,\theta,t-1) + \left[{{S_{u}(r-1,\theta,t)\, \Delta t} \over {{4 \pi\, n_{\rm csm}(r,\theta)\, r_{\rm WD}^2\, \Delta r\, f_{\rm vol}(\theta)}}}\right] \\
& - \left[{{4 \pi\, x(r,\theta,t-1)^2\, n_{\rm csm}(r,\theta)^2\, \alpha_{(2)}\, r_{\rm WD}^2\, \Delta r\, f_{\rm vol}(\theta)\, \Delta t} \over {{4 \pi\, n_{\rm csm}(r,\theta)\, r_{\rm WD}^2\, \Delta r\, f_{\rm vol}(\theta)}}}\right] 
\end{split}
\end{equation}
This formulation properly accounts for those time steps and bins which are not completely neutral ($x > 0$) and are not reached by any ionizing flux; these will slowly recombine in the absence of an ionizing radiation field.

\subsubsection{Nova Shock Front Model}\label{shock_model}
Because $10^{7}$ K ionized gas has very different radio-emitting properties than $10^{4}$ K gas, we must also track the location of the nova shock as a function of time and $\theta$. Increasing the electron temperature by three orders of magnitude decreases the radio optical depth by four orders of magnitude (Equation \ref{eq:tau}). The nova blastwave will sweep up circumbinary material and heat it from radio-emitting temperatures to X-ray emitting temperatures. 

Our model of the evolving nova blastwave is as described in \citet{Nelson_etal12}. We input $\dot{M}_w$, $v_w$, $a$, $M_{\rm ej}$, and initial blast wave velocity. The model assumes that the ejecta are expanding in a thin shell at a single velocity initially. Each polar-angle ring then has an independent evolution of radius and velocity that depends on the density profile it encounters. It decelerates as it sweeps up material, conserving kinetic energy in a Sedov-like evolution.

\begin{figure*}
\centering
\includegraphics[width=16cm, angle=90]{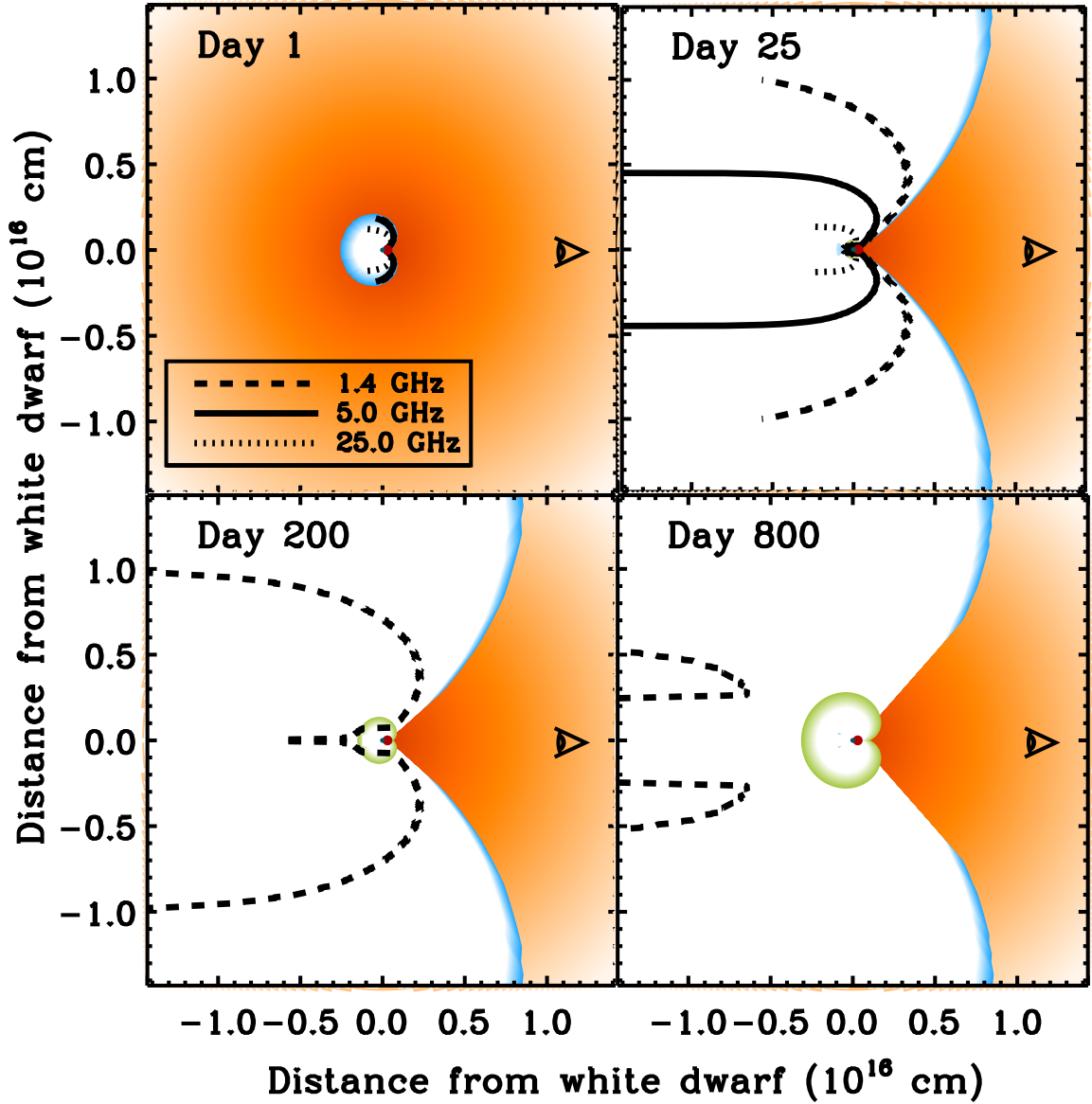}
\caption{Models of V407~Cyg, with the binary located at the origin (small red and blue dots; not to scale) and the nova expanding into a Mira wind of mass-loss rate $\dot{M}_w = 2 \times 10^{-6}$ M$_{\odot}$ yr$^{-1}$, wind velocity $v_w = 10$ km s$^{-1}$, and wind temperature $T_{e,w} = 1 \times 10^{4}$ K (orange gradient), as pictured in Figure \ref{ill}. We show the environment of the binary at four different epochs: 1, 25, 200, and 800 days after outburst. In each epoch, the portion of the wind that is ionized is filled with white and outlined in blue. The extent of the nova shock front is outlined in green (too small to be visible here on Days 1 and 25). Surfaces where optical depth $\tau = 1$ is reached are shown as black lines, and are depicted at three different frequencies: 1.4 GHz (dashed), 5.0 GHz (solid), and 25.0 GHz (dotted). These surfaces grow as the wind becomes more ionized, but are destroyed when the nova shock sweeps over them.}
\label{ppt}
\end{figure*}

\subsubsection{Calculation of Radio Flux Densities}

The luminosity of thermal radio emission depends critically on the emission measure (EM), here defined as $\int n_{e}^2\,dl$, where $n_{e}$ is the electron density of the absorbing material and $dl$ is the path length through this material. The optical depth $\tau_{\nu}$ at an observing frequency $\nu$ along a given line of sight is:
\begin{equation}\label{eq:tau}
\tau_{\nu} = 8.235 \times 10^{-2}\, \,\left({{T_e}\over{\rm K}}\right)^{-1.35}\,\left({{\nu}\over{\rm GHz}}\right)^{-2.1}\, \left({{\rm EM}\over{\rm cm^{-6}\ pc}}\right) 
\end{equation}
\citep{Seaquist_Bode08}. The flux density $S_{\nu}$  produced by this line of sight is:
\begin{equation}
\begin{split}
\left({{S_{\nu}}\over{\rm mJy}}\right) = 3.2 \times 10^{-36}\, &\left({{A}\over{\rm cm^{-2}}}\right)\, \left({{D}\over{\rm kpc}}\right)^{-2}\,\left({{T_e}\over{\rm K}}\right) \times \\
&\left({{\nu}\over{\rm GHz}}\right)^{2} \left(1 - e^{-\tau_{\nu}}\right).
\end{split}
\end{equation}
Here, $A$ is the projected area of this line of sight, $D$ is the distance to the emitting source, and $T_{e}$ is its electron temperature.

\begin{figure}
\centering
\includegraphics[width=11.6cm, angle=90]{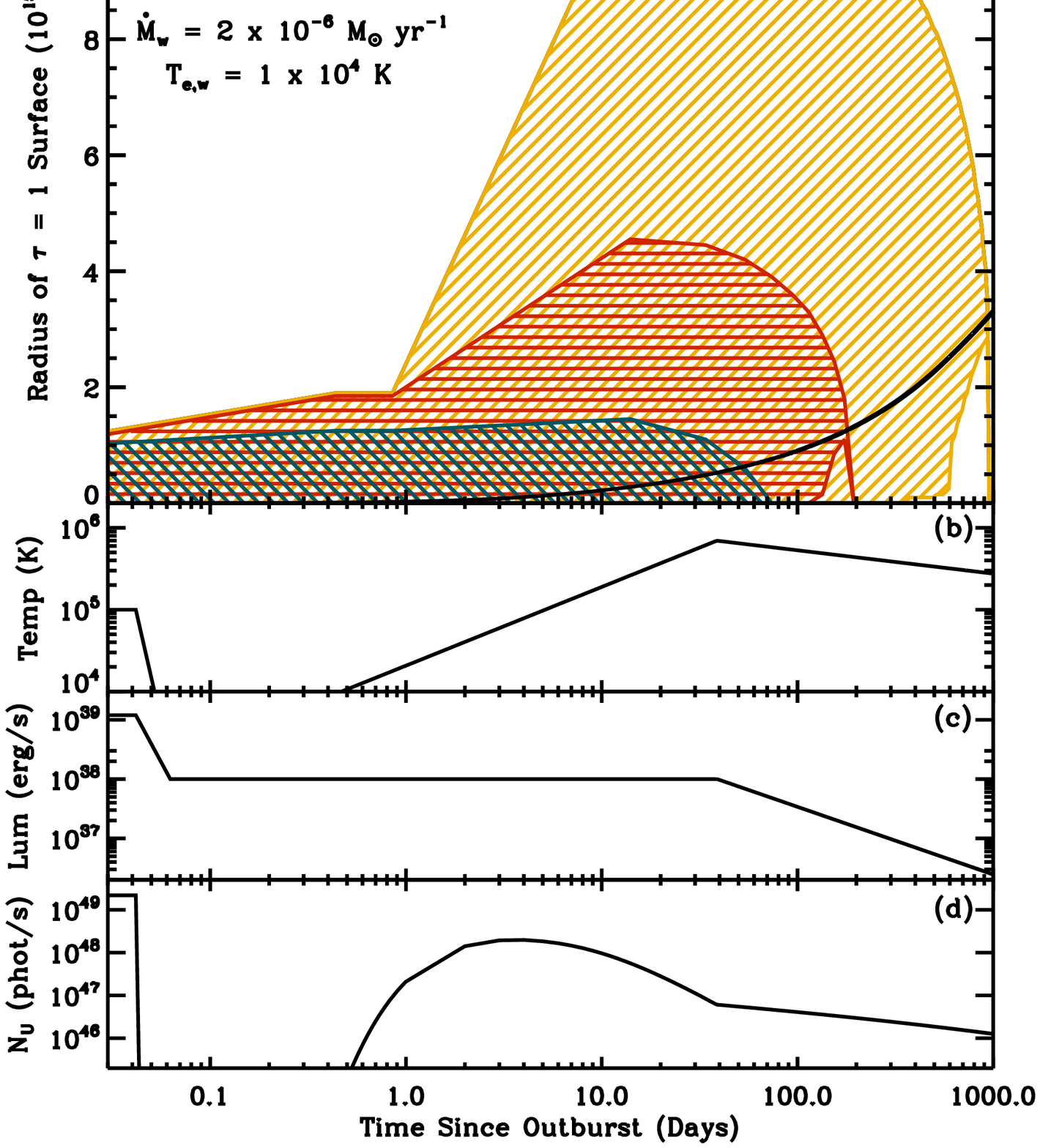}
\caption{\emph{Panel (a):} For the same model parameters as depicted in Figure \ref{ppt}, we plot the radius of the $\tau = 1$ surfaces, perpendicular to our line of sight, as a function of time for 1.4 GHz (gold), 5.0 GHz (red), and 25.0 GHz (blue).The maximum radius of the nova blastwave is also shown as a solid black line. \emph{Panel (b):} The assumed evolution of the WD blackbody temperature. \emph{Panel (c):} The assumed evolution of the WD blackbody luminosity. \emph{Panel (d):} The ionizing photon flux produced by the assumed WD temperature and luminosity evolution.}
\label{taurad}
\end{figure}

In our model of V407~Cyg described above, we can calculate $S_{\nu}$ if we assume a viewing angle; optical spectroscopy observationally constrains this angle for the 2010 outburst of V407~Cyg. \citet{Munari_etal11} and \citet{Shore_etal12} point out that narrow absorption lines of hydrogen and metals are present in the spectrum of V407~Cyg following the nova outburst, implying the presence of neutral circumstellar material. Although these lines fade with time, they never completely disappear, implying that the wind never becomes completely ionized. Therefore, the nova was probably observed while the giant was in front of the WD along our line of sight \citep{Shore_etal12}. Our best guess at the orbital period of V407~Cyg (43 years; \citealt{Munari_etal90}) is significantly longer than the timescale of our observations, so we assume this viewing angle is constant across our radio light curve. We show a schematic picture of the binary in Figure \ref{ill}, which we assume is being viewed from the right-hand side of the figure.

While our model is not currently equipped to provide light curves for an arbitrary viewing angle, the radio light curves presented here should be relatively unaffected by small changes in viewing angle, as the radio luminosity is predominately dependent on the projected angular area of warm ionized wind, which will weakly depend on viewing angle. This insensitivity to viewing angle is illustrated for symbiotic stars in Figure 6 of \citet{Seaquist_etal84}, where, for winds that are significantly ionized, the measured flux densities and spectral indices change negligibly with a 90$^{\circ}$ change in viewing angle.

We assume an electron temperature and construct an array of impact parameters (values of $z$, separated by $\Delta z$), along the axis perpendicular to our line of sight (Figure \ref{ill}). For a given impact parameter, we find all bins of $(r, \theta)$ that are intercepted by this line of sight, and sum them up. The EM for each bin is calculated as:
\begin{equation}
\begin{split}
{\rm EM}&(r,\theta,t) = {{x(r,\theta,t)^2\  n_{\rm csm}(r,\theta)^2} \over {3.086 \times 10^{18}}} \times \\
& \left[{{z}\over{{\rm tan}(\theta + \frac{\Delta \theta}{2})}} - {{z}\over{{\rm tan}(\theta - \frac{\Delta \theta}{2})}}\right]\, {\rm cm}^{-6}\ {\rm pc}.
\end{split}
\end{equation}
However, a bin is only considered as contributing to the radio-emitting EM if it has not been shocked by the nova blastwave. If a bin has been swept over by the nova shock, calculated as described in Section \ref{shock_model} and \citet{Nelson_etal12}, it will typically be heated to $10^{6}-10^{8}$ K, and will therefore be optically thin and faint at radio wavelengths; we set EM$(r, \theta, t) = 0$ in this case. We calculate $A$ by assuming that the emitting area at a given impact parameter is ring-shaped: $A = \pi ((z+ \frac{\Delta z}{2})^2 - (z- \frac{\Delta z}{2})^2)$.

To understand the evolution of the optically thick radio emission, we find it instructive to calculate surfaces where the optical depth reaches a value of unity, as a function of time. Radio observations at a given frequency $\nu$ will be insensitive to conditions within the $\tau_{\nu} = 1$ surface. The size of a $\tau_{\nu} = 1$ surface depends on the path length through and density of the ionized gas. To reach $\tau_{\nu} = 1$, an EM of ionized material must be obtained equivalent to:
\begin{equation}\label{em}
\left({{\rm EM}_{\tau}\over{\rm cm^{-6}\ pc}}\right) = 12.1\,\left({{T_e}\over{\rm K}}\right)^{1.35}\,\left({{\nu}\over{\rm GHz}}\right)^{2.1}.
\end{equation}
To calculate the shape and size of the $\tau_{\nu} = 1$ surface at a given time and frequency, we again step through values of $z$. At a given impact parameter, we proceed from the bin closest to us along our line of sight, located far out in the Mira wind, inward toward the binary and potentially beyond it, summing up the EM in bins until EM$_{\tau}$ is reached. The bin where EM$_{\tau}$ is obtained marks a location in $(r, \theta)$ of the $\tau = 1$ surface. This process is iterated for all impact parameters to obtain an array of such locations and map the topology of the $\tau = 1$ surface (plotted as black lines in Figure \ref{ppt}). 

In Figure \ref{taurad}, we plot the evolution of the radius of the $\tau = 1$ surface at three different frequencies, measured perpendicular to our line of sight. The radius of the $\tau = 1$ surface is defined as the maximum impact parameter where $\tau = 1$ is reached. In some models, the $\tau = 1$ surface does not always look like a filled circle, but displays a ring-like geometry at later times, because EM$_{\tau}$ is never reached at small values of $z$ (e.g., Day 800 in Figure \ref{ppt}). Ring-shaped $\tau = 1$ surfaces are produced by the nova blast eating away at the thermal wind from the inside. They are depicted in Figure \ref{taurad} as having both a minimum and maximum radius (e.g., the 1.4 GHz curve in Figure \ref{taurad}, between Days 500 and 1000).

\subsubsection{Assumed Physical Parameters} \label{ionass}

In exploring the parameter space of V407~Cyg, we hold most parameters fixed and focus on varying the wind mass loss rate, $\dot{M}_{w}$. The wind velocity is assumed to be 10 km s$^{-1}$ in our discussion of $\dot{M}_{w}$. The binary separation is held constant at 17 AU. In calculating the evolution of the nova blastwave, we assume an initial shock velocity of 3,200 km s$^{-1}$ \citep{Munari_etal11, Shore_etal11} and an ejecta mass of $10^{-6}$ M$_{\odot}$.

\begin{figure}
\centering
\includegraphics[width=9cm, angle=90]{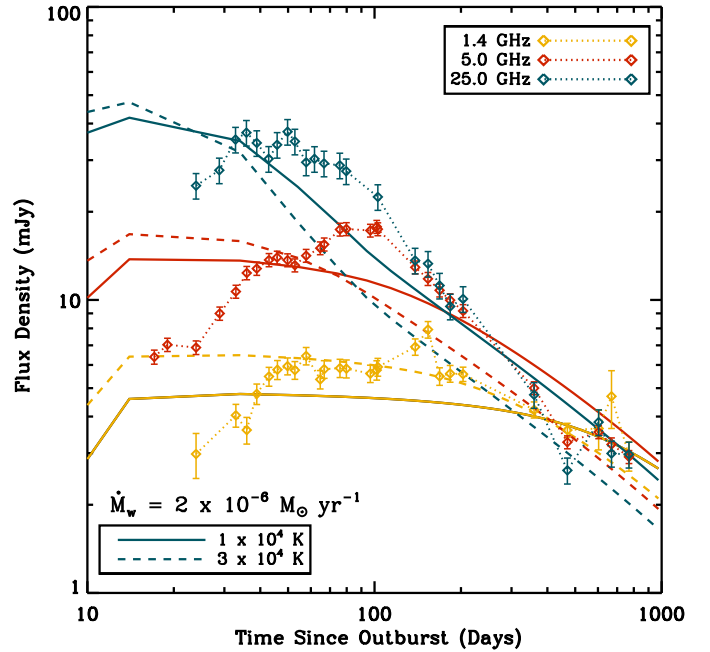}
\caption{Light curves for V407~Cyg at three frequencies: 1.4 GHz (gold), 5.0 GHz (red), and 25.0 Ghz (blue). The diamond-shaped points connected by dotted lines are our VLA data. Models assuming $\dot{M}_w = 2 \times 10^{-6}$ M$_{\odot}$ yr$^{-1}$ and $T_{e,w} = 1 \times 10^{4}$ K (as in Figure \ref{ppt}) are shown as solid lines. Dashed lines are models for $\dot{M}_w = 2 \times 10^{-6}$ M$_{\odot}$ yr$^{-1}$ and $T_{e,w} = 3 \times 10^{4}$ K.}
\label{modgood}
\end{figure}

We assume a wind electron temperature of $T_{e,w} = 1 \times 10^{4}$ K, but the calculated radio emission is relatively insensitive to $T_{e,w}$. In Figure \ref{modgood}, we plot the same model, one with  $T_{e,w} = 1 \times 10^{4}$ K (solid lines) and the other with  $T_{e,w} = 3 \times 10^{4}$ K (dashed lines). At most points in the light curve, the models differ by less than a factor of $\sim$2.

An essential input for our models is the ionizing flux that emanates from the nuclear-burning WD, and how this flux varies with time. We assume the WD has a blackbody spectrum; theory and X-ray observations of V407~Cyg inform our assumptions of how the blackbody temperature and luminosity vary with time. We plot the assumed evolution of the hot WD in the bottom three panels of Figure~\ref{taurad}.

At the onset of the thermonuclear runaway, a brief ``flash" of ionizing radiation is released, before the ejecta has time to expand and reprocess the hot WD spectrum to a puffed-up $\sim$10,000 K blackbody. We assume this flash lasts 1 hour, has a luminosity of $1.2 \times 10^{39}$ erg s$^{-1}$, and a temperature of $10^{5}$ K \citep{Prialnik86}. Subsequently, the ionizing flux dramatically diminishes, until the ejecta have expanded sufficiently to become optically thin to the hot WD once more.

There is evidence for significant ionizing flux in the \emph{Swift} X-ray light curve from a soft X-ray component present between Days $\sim$15--50 \citep{Shore_etal11, Schwarz_etal11, Nelson_etal12}. The soft X-ray component is likely attributable to the hot WD, which would then be the primary source of ionizing flux in the V407~Cyg 2010 nova. Note that the hard X-ray component (attributed to forward-shocked material in the nova blastwave) is observed to be brighter than the soft X-ray component in the \emph{Swift lightcurve}, but this is only an observational effect produced by the severe absorption of the soft X-rays by circumstellar material. Because of the strong absorption, it is difficult to precisely determine the luminosity of the hot WD from the \emph{Swift} data \citep{Schwarz_etal11, Nelson_etal12}. We therefore assume a WD plateau luminosity of $1 \times 10^{38}$ erg s$^{-1}$, as consistently implied by models of nuclear burning on WDs during nova events \citep[e.g.,][]{Sala_Hernanz05}. We do use the \emph{Swift} measurements to constrain the rising temperature of the WD up until a peak temperature on Day 39. The rise is fit roughly with $T_{\rm WD} = 2 \times 10^{4}\, t^{0.96}$ K, where $t$ is the time since outburst in units of days. Note that this fit is extrapolated back to the time of onset of outburst, and implies that the ionizing flux from the WD is peaking just a few days after outburst (Figure \ref{taurad}). This early-time evolution is highly uncertain and unconstrained by data, and yet unfortunately it is the primary determinant of the early time radio light curve. After the temperature peak on Day 39, we assume the luminosity of the WD declines from its plateau value as $L_{\rm WD} \propto t^{-1.14}$, with a roughly constant radius so $T_{\rm WD} \propto t^{-0.29}$ \citep{Prialnik86}.

\subsection{Interpreting the Light Curve with Our Model}

The ionization state of the Mira wind varies in the aftermath of the nova explosion, as evidenced by optical spectroscopy. Narrow H$\alpha$ emission lines are seen at early times (just two days after outburst), consistent with an ionized wind \citep{Munari_etal11}. 
%Throughout the entire evolution of the outburst, there was a narrow H$\alpha$ absorption line superimposed on the H$\alpha$ emission \citep{Munari_etal11}, implying that some neutral material remains in the Mira wind. 
In the first months after explosion, narrow P~Cygni profiles were observed for low-ionization metal lines \citep{Shore_etal12}. These P~Cygni metal lines disappeared by Day 70, while the Balmer line absorption weakened over this period \citep{Shore_etal12}. Shore et al.~attribute this to increasing ionization in the Mira wind over the months following outburst. This increase in ionization of the wind is not surprising, given observations of a hot WD between Days $\sim15-50$ (Section \ref{ionass}). It is worth noting that some H$\alpha$ absorption was sustained at all times, implying that the wind never becomes completely ionized \citep{Munari_etal11}. 

In our VLA data, the ionized wind of the Mira giant radiates as free-free emission, just as detected for other symbiotic stars in quiescence \citep[e.g.,][]{Seaquist_Taylor90}. The spectral index at early times ($\alpha = 0.6-1.0$) is consistent with the theoretical expectation for a spherical $\rho \propto r^{-2}$ wind \citep[$\alpha = 0.6$;][]{Olnon75, Panagia_Felli75, Wright_Barlow75} and spectral index measurements for symbiotic stars ($\alpha$ ranging from $-0.1$ to 1.4, with a median of 0.8; \citealt{Seaquist_Taylor90}). However, V407~Cyg is more complex than the typical symbiotic star, because the radio flux densities are changing rapidly with time. Our model shows that the 1.4--25 GHz light curves rise at early times because the wind becomes more ionized (Figures \ref{ppt} and \ref{modgood}). In other words, the $X$ parameter, as defined in \citet{Seaquist_etal84} to parameterize the fractional ionization of the wind, is increasing with time. Seaquist et al.~show that as $X$ increases, the spectral index of the partially optically-thick emission flattens---consistent with the decrease in $\alpha$ from 1.0 to 0.7 between Days 12 and 39. 

As $X$ increases and the ``cone" of neutral wind decreases in extent, the ionized emission measure at the outer extents of the wind increases. The radio flux density thereby brightens and the $\tau_{\nu} = 1$ surfaces grow in diameter (e.g., compare Days 1 and 25 in Figure \ref{ppt}). We stress that our radio light curve can not probe conditions internal to the $\tau_{\nu} \approx 1$ surface. These surfaces are at radii of $\gtrsim 10^{15}$ cm, significantly larger than the nova blastwave radius (black solid line in the top panel of Figure \ref{taurad}). Complex processes may be occurring within the $\tau_{\nu} = 1$ surfaces, like thermal emission from expanding nova ejecta (Section \ref{ejecta}) or synchrotron emission from blastwave interaction (Section \ref{synch}). However, this emission would only become apparent when either the wind becomes optically thin or the additional emitting components expand to radii larger than the $\tau_{\nu} \approx 1$ surface.

As can be seen in Figure \ref{modgood}, the model agrees with our observed radio light curves as well as can be expected, given the significant uncertainties in the ionizing source. The model plotted in Figure \ref{modgood} assumes a Mira wind mass-loss rate of $\dot{M}_{w} = 2 \times 10^{-6}$ M$_{\odot}$ yr$^{-1}$, a distance of 3 kpc, and $T_{e,w} = 10^{4}$ K. It broadly reproduces the observed features of the radio light curve. The peak flux densities are roughly matched (to within a factor of $\sim$2; compare this with the order-of-magnitude discrepancy at lower frequencies in Figure \ref{lcmod}), as are the slow decline times and the earlier turn-over at higher frequency. 

However, the model light curves peak too early, on Day $\sim$15, relative to the observed maxima of Day $\sim30-100$. Revising the input ionizing radiation field and its evolution---a tweak completely justified given the paucity of observations constraining the hot WD at early times---would ease this discrepancy. We have no constraints on the WD's luminosity or temperature before Day 15 or after Day 50, while its spectrum is peaking in the UV rather than the X-ray---even though the WD is likely an important ionizing source at these times. We assume that the increase in temperature and luminosity of the WD is gradual; extrapolating from the observed temperature and luminosity points implies that the ionizing flux is actually peaking very early, around Day 4 (bottom panel of Figure \ref{taurad}). If, instead, the temperature and luminosity of the WD increased suddenly, as observed in RS Oph by \citet{Osborne_etal11}, the predicted rise of the radio light curve would change significantly. More detailed modeling of the luminosity and temperature evolution of the WD is outside the scope of this paper.

\begin{figure}
\centering
\includegraphics[width=9cm]{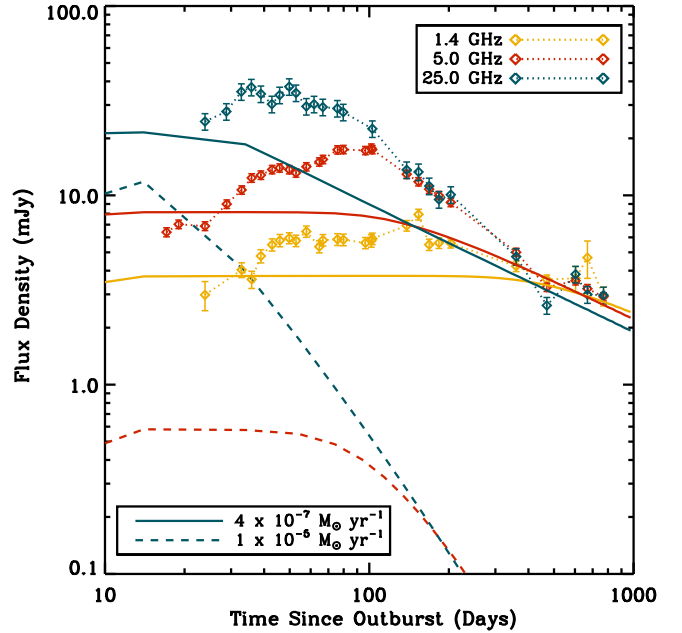}
\caption{Light curves for V407~Cyg; similar to Figure \ref{modgood} but showing models with different $\dot{M}_w$. Diamond-shaped points connected by dotted lines are our VLA data. Solid lines show a model with $\dot{M}_w = 4 \times 10^{-7}$ M$_{\odot}$ yr$^{-1}$. Dashed lines are models for $\dot{M}_w = 1 \times 10^{-5}$ M$_{\odot}$ yr$^{-1}$; the 1.4 GHz model is so faint that it falls outside the plot limits. Both sets of model lines assume $T_{e,w} = 1 \times 10^{4}$ K.}
\label{modbad}
\end{figure}

If the mass-loss rate is significantly lower than $\dot{M}_{w} = 2 \times 10^{-6}$ M$_{\odot}$ yr$^{-1}$ (solid lines in Figure \ref{modbad}), the flux densities never become large enough to match our observations, even if the entire wind is ionized. We can understand this by applying the standard equation for deriving mass-loss rates of fully-ionized stellar winds from thermal radio flux densities \citep{Wright_Barlow75}:
\begin{equation}\label{eq:mdot}
\begin{split}
\dot{M}_w =&\ 4.5 \times 10^{-7}\  \left({{v_w}\over{10\, \textrm{km s}^{-1}}}\right)\ \left({{S_{\nu}}\over{1\, {\rm mJy}}}\right)^{3/4} \times \\
&  \left({{D}\over{3.0\,{\rm kpc}}}\right)^{3/2} \left({{\nu}\over{1\,{\rm GHz}}}\right)^{-1/2}\ g_{ff}^{-1/2}\ {\rm M}_{\odot}\ {\rm yr}^{-1}.
\end{split}
\end{equation} 
The Gaunt factor $g_{ff}$ is a weak function of the wind electron temperature:
\begin{equation}
g_{ff} = 9.77\ \left(-0.17 + 0.130\,{\rm log}{T_{e,w}^{3/2}\over{\nu}}\right).
\end{equation} 
We estimate $S_{\nu}$ from the broad maxima of our 1.4--25 GHz light curves, and find $\dot{M}_w \approx (6.6-7.7) \times 10^{-7}~{{\rm M}_{\odot}\ {\rm yr}^{-1}}$ for a distance of 3.0 kpc and a wind electron temperature of $T_{e,w} = 10^4$ K. This estimate is only very weakly dependent on the wind temperature. For $T_{e,w} = 30,000$ K, $\dot{M}_w$ decreases slightly to $(6.0-7.1)\times 10^{-7}~{{\rm M}_{\odot}\ {\rm yr}^{-1}}$. These are essentially lower limits on $\dot{M}_w$. Of course, if V407~Cyg is at a smaller distance, the implied $\dot{M}_w$ will drop accordingly ($\dot{M}_w \approx 2.5 \times 10^{-7}~{{\rm M}_{\odot}\ {\rm yr}^{-1}}$ at 1.5 kpc).

If the mass-loss rate is significantly higher than $\dot{M}_{w} = 2 \times 10^{-6}$ M$_{\odot}$ yr$^{-1}$, as shown by the dashed-line model in Figure \ref{modbad}, the hot WD will only be able to ionize a relatively small volume. As a result, like in the low $\dot{M}_w$ case, we predict lower-than-observed flux densities and small $\tau = 1$ surfaces (although, in this case the flux density will depend rather sensitively on the assumed luminosity and evolution of the ionizing radiation field). 

Not only do small $\tau = 1$ surfaces produce too little radio luminosity, they also lead to radio light curves that decline too quickly, because the nova shock front sweeps over them at earlier times. For example, in the high $\dot{M}_{w}$ case plotted in Figure \ref{modbad}, the model predicts that the radio emission at 25 GHz should be optically thin around the time the observed light curve is peaking. 

Radio flux densities eventually decline for two reasons: recombination and heating by the nova shock. A good example of decline by recombination can be seen in the 1.4 GHz curve in Figure \ref{taurad}; the maximum radius declines from $\sim 11 \times 10^{15}$ cm to $\sim 4 \times 10^{15}$ because of recombination in the outer parts of the wind. At the same time, between Days 600 and 1000, we see that the minimum radius of the 1.4 GHz $\tau = 1$ surface increases from 0 cm to $\sim 4 \times 10^{15}$ cm, until the minimum and maximum radii meet and the system becomes completely optically thin to 1.4 GHz radiation. 

In our observed light curves, all frequencies stop rising after Day $\sim$70---consistent with the findings of \citet{Shore_etal12} that the wind becomes increasingly ionized up through Day $\sim$70. Soon thereafter, the observed light curves turn over and begin to decline for the higher frequencies ($>$4 GHz), but the 1.4/1.8 GHz light curves are constant for another $\sim$100 days before declining and becoming optically thin. Correspondingly, our model 1.4 GHz curve is quite flat-topped until Day $\sim$300, when both recombination and the nova blast begin to significantly diminish the radio emission measure. 

At late times, the radio flux densities observed at all frequencies converge to a steady (or perhaps slowly declining) $\sim$3.5 mJy (starting around Day 350, continuing through the time of submission). The radio spectrum is consistent with optically-thin thermal emission ($\alpha \approx -0.1$; Table \ref{tab:spind}). Such a leveling-off of the radio light curves is, to date, consistent with our rough model for the recombining Mira wind (Figure \ref{modgood}). Occasional monitoring of V407~Cyg in the upcoming years will show if the radio luminosity continues to gradually fade as expected.

\section{Why No Radio Synchrotron Emission Observed in V407~Cyg?} \label{synch}

\begin{figure*}
\centering
\includegraphics[width=11cm, angle=90]{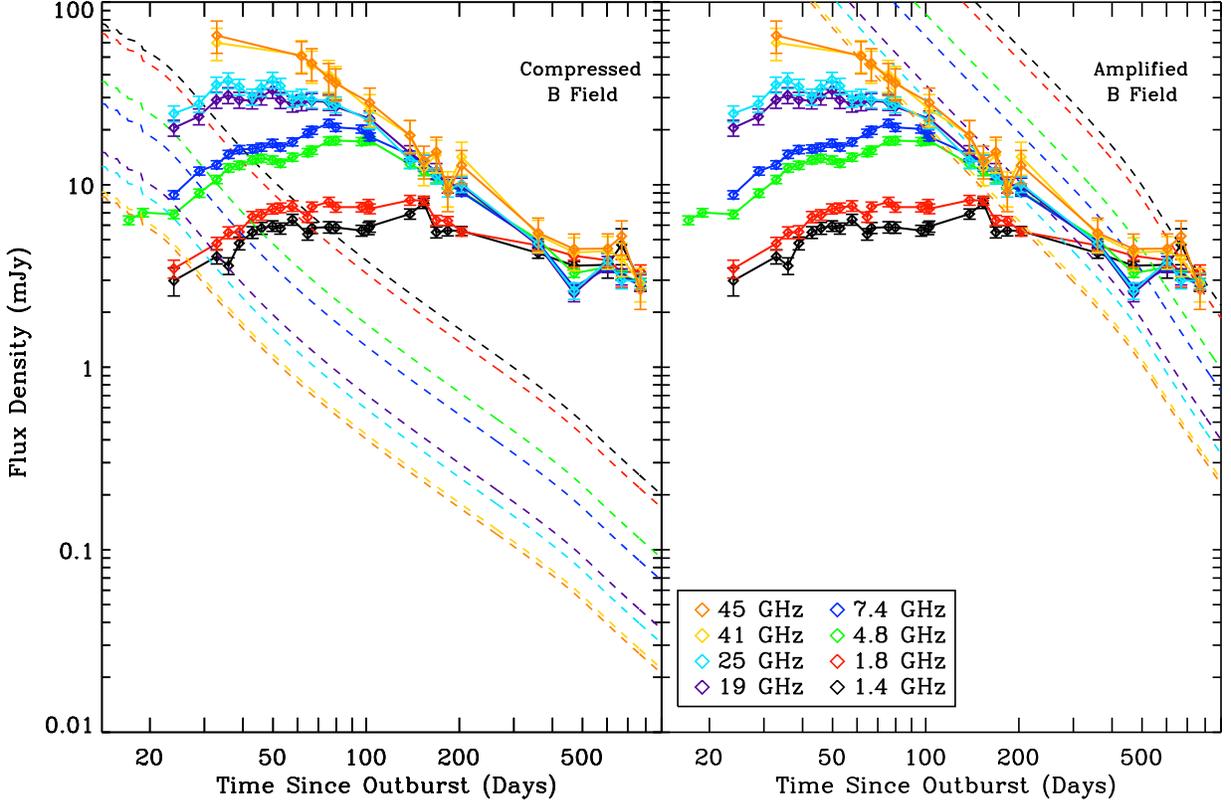}
\caption{Models for synchrotron emission from V407~Cyg, overplotted as dashed lines on our measured multi-frequency light curve (solid lines and diamonds). The two panels represent different assumptions about the magnetic field strength in the forward-shocked region: the left panel assumes $B_{\rm comp}$ as described in Equation \ref{bcomp}, while the right panel assume $B_{\rm amp}$ from Equation \ref{bamp}.}
\label{lcsynch}
\end{figure*}

From the gamma-ray detection of V407~Cyg, we know that particles were accelerated to relativistic speeds by the nova shock \citep{Abdo_etal10}. We might therefore expect accompanying radio synchrotron emission. However, at no time are our measured spectral indices (Table \ref{tab:spind}) indicative of optically-thin synchrotron emission, for which we would expect $\alpha = -0.5$ to $-1.0$. Here we model the synchrotron emission from V407~Cyg to show that bright synchrotron emission was likely produced in V407~Cyg, but it was hidden behind the absorbing screen of the Mira wind.

We assume that as the nova shock wave interacts with surrounding circumstellar material, it converts a constant fraction ($\epsilon$) of the post-shock energy density into relativistic electrons, such that:
\begin{equation}
\int_{\rm 0.51\ MeV}^{\infty} N_0\ E^{-p}\ dE = \epsilon\, \mu\, m_H\, n_{\rm CSM}\, v_s^{2},
\end{equation}
where all quantities are in cgs units. The blastwave velocity is $v_s$ and the Mira wind density at the location of the blastwave is $n_{\rm CSM}$ (as described in Equation \ref{eq:dens}). Here, $p$ is the index of the electron energy spectrum, assumed to have a power-law form. \citet{Abdo_etal10} estimate $p$ with large uncertainties from the gamma-ray spectrum: $p = 1.75^{+2.4}_{-0.6}$. We can also directly measure $p$ if optically-thin synchrotron emission is detected, as it is in a few other novae. The radio spectral index implies $p = 2.1-2.5$ in RS Oph \citep{Rupen_etal08, Sokoloski_etal08} and $p=2.4$ in GK Persei \citep{Seaquist_etal89}; we assume $p=2.3$ for V407~Cyg. The minimum energy in the integral, 0.51 MeV, corresponds to the rest mass energy of the electron and approximates the energy at which electrons become relativistic. The efficiency of particle acceleration, $\epsilon$, is assumed to be constant with time. \citet{Abdo_etal10} show that, if the gamma-ray emission was solely produced by leptonic processes, then $\epsilon \approx 0.004$; if hadronic processes contribute to the gamma-ray luminosity, this is an upper limit.
 
The synchrotron luminosity also depends on the magnetic field strength, and we consider two physical scenarios for its strength and evolution. In the first case, we assume that the magnetic field in the forward-shock region is simply the magnetic field from the Mira wind, compressed. As suggested by \citet{Abdo_etal10}, the magnetic field energy density in the wind is expected to be in equipartition with the thermal energy density of the wind. We assume that the relevant conditions pertain to the wind before it was heated and ionized by the nova outburst, so we set the wind temperature to $T_{w0} = 700$ K as in Abdo et al. The magnetic field strength is then simply a function of distance from the Mira giant, and is described as:
\begin{equation}\label{bcomp}
B_{\rm comp} = \sqrt{{{32\, \pi\, n_{\rm CSM}\, k\, T_{w0}}}}.
\end{equation}

Another possibility is that the magnetic field is amplified in the post-shock region to be in equipartition with the relativistic electrons. This simplification is commonly used to interpret the radio emission from SNe \citep[e.g.,][]{Chevalier_Fransson06}, and was recently utilized by \citet{Chugai10} for outbursts in symbiotic stars. In this case:
\begin{equation}\label{bamp}
B_{\rm amp} = \sqrt{8 \pi\,\epsilon\,\mu\, m_H\, n_{\rm CSM}\,v_s^{2}}.
\end{equation}
Here, $B$ depends on both the distance from the Mira giant and the blastwave velocity.

The synchrotron flux density of a spherical body of radius $R$ at distance $D$ (where both $R$ and $D$ are in units of cm) can be found by Equation 1 of \citet{Chevalier98} to be:
\begin{equation}\label{snu_synch}
\begin{split}
S_{\nu} &= 5.64 \times 10^{-22}\ {{c_5}\over{c_6}}\ B^{-1/2}\ \nu^{5/2} \times \\
& \left(1 - {\rm exp}\left[-{{\nu}\over{\nu_1}}^{-(p+4)/2}\right]\right)\ \left(R/D\right)^2\ \  {\rm mJy}
\end{split}
\end{equation}
Here, $c_5$ and $c_6$ are constants that depend on $p$ as tabulated in \citet{Pacholczyk70}, and as listed in Table \ref{tab:pach} for convenience. The observing frequency is $\nu$ in units of Hz, and $\nu_1$ is defined as:
\begin{equation} \label{nu1}
\nu_{1} = 1.25 \times 10^{19}\ \left({{4}\over{3}}\,f\,R\,c_6\,N_0\right)^{2/(p+4)}\ B^{(p+2)/(p+4)}
\end{equation}
We assume that the synchrotron emission is contained in a spherical shell between the forward shock and the contact discontinuity, typically with a width $\sim$20\% of the contact discontinuity's radius \citep{Chevalier82a}. In Equation \ref{nu1}, $f$ is the volume filling factor of this emission, so $f = 0.42$ for a width of 20\%.

All the above equations assume spherical symmetry, but we know that V407~Cyg is an inherently asymmetric case. Because synchrotron luminosity is sensitive to both shock velocity and circumstellar density, it is important that we take into account this asymmetry in predicting the expected synchrotron signal from V407~Cyg. We again use the toy model of \citet{Nelson_etal12} to estimate the evolution of the blastwave in this asymmetric medium (see Section \ref{shock_model}). We calculate the synchrotron flux density emanating from each polar-angle ring by using Equation \ref{snu_synch} and scaling the flux density by the fraction of a spherical shell's volume that is enclosed in that ring ($f_{\rm vol}$; Equation \ref{fvol}). The flux densities from all rings are summed together to give the total synchrotron flux at a given time, producing the model light curve plotted in Figure \ref{lcsynch}. 

The light curves in Figure \ref{lcsynch} show the evolution of this synchrotron model, assuming $M_{\rm ej} = 10^{-6}$ M$_{\odot}$, $\dot{M}_w = 2 \times 10^{-6}$ M$_{\odot}$ yr$^{-1}$, $v_w = 10$ km s$^{-1}$, $\epsilon = 0.004$, $a = 17$ AU, an initial shock velocity of $v_{s} = 3,200$ km s$^{-1}$, and a distance of 3.0 kpc. The left panel assumes a compressed magnetic field (Equation \ref{bcomp}), while the right panel assumes an amplified magnetic field in equipartition with the relativistic electrons (Equation \ref{bamp}). The synchrotron emission in both models is optically thin, with spectral index $\alpha = -0.65$, so the lower frequencies are brighter. The synchrotron flux density decreases with time because the blastwave is decelerating and the density of the circumstellar material is declining in all directions that are significantly contributing to the synchrotron emission. The predicted synchrotron emission in the $B_{\rm comp}$ case is much less luminous than in the $B_{\rm amp}$ case, because the magnetic field in the forward shock is weaker. 

Both models predict bright synchrotron emission that would have been easily detectable with the VLA in the first months after outburst, with flux densities $>> 1$ mJy (a result confirmed by the models of \citealt{Martin_Dubus12}). However, in reality at early times, when the synchrotron luminosity was highest, the synchrotron-emitting forward shock suffered free-free absorption from the ionized Mira wind (compare the black solid line in Figure \ref{taurad} with the radii of $\tau = 1$ surfaces). The large radius of the absorbing screen means synchrotron emission is only visible at a given frequency once the wind becomes optically thin at that frequency. A comparison between synchrotron models and observations is most simply carried out on Day $\sim$470, when the observed emission from V407~Cyg is consistent with optically thin thermal radiation at all observed frequencies. In the $B_{\rm comp}$ model, synchrotron emission constitutes a trivial fraction of the total flux density on Day 470, even at 1.4 GHz. In the $B_{\rm amp}$ model, on the other hand, the predicted synchrotron luminosities at lower frequencies ($\lesssim8$ GHz) are significantly higher than those observed on Day 470, and we might have expected to see a clear signature of synchrotron emission in the spectral evolution. We stress that no such synchrotron contribution is observed in our radio spectra; spectral indices measured between Days 359--669 are $\alpha \approx 0$, as expected for optically-thin thermal emission ($\alpha = -0.1$) and inconsistent with expectations for synchrotron emission ($\alpha \approx -0.65$). We may be able to constrain the magnetic field in the forward shock with our radio light curve.

Of course, our model of the synchrotron emission depends quite sensitively on a number of assumed physical parameters. Although V407~Cyg is a better-constrained system than many, thanks to the community's wealth of multi-wavelength data, there is still room to vary predictions of the synchrotron luminosity by orders of magnitude. For example, a decrease of the ejecta mass from $M_{\rm ej} = 10^{-6}$ M$_{\odot}$ to $M_{\rm ej} = 10^{-7}$ M$_{\odot}$ means that the blast wave decelerates much faster, and decreases the 1.4 GHz flux density predicted by the $B_{\rm amp}$ model on Day 470 from 11.5 mJy to 0.2 mJy. Because synchrotron luminosity depends so sensitively on blast wave velocity, we are likely too simplistic in assuming that the ejecta are expelled in a thin shell with a single velocity; the synchrotron model might be improved with more realistic radial distributions of ejecta mass and velocity. The $B_{\rm amp}$ model depends on shock velocity more sensitively than the $B_{\rm comp}$ model; decreasing $M_{\rm ej}$ to $10^{-7}$ M$_{\odot}$ in the $B_{\rm comp}$ case only decreases the 1.4 GHz flux density from 0.55 mJy to 0.11 mJy. In the $B_{\rm amp}$ case, the energy densities in both the magnetic field and relativistic electrons depend on the shock velocity, while in the $B_{\rm comp}$ case, the magnetic field is independent of shock velocity. 

The synchrotron luminosity also depends sensitively on assumptions about the efficiencies of particle acceleration and $B$ field amplification in the shock. Although V407~Cyg is unusual because $\epsilon$ has been directly constrained from the gamma-ray observations, this constraint remains a rather uncertain upper limit. Additional uncertainties are involved in our assumptions about the evolution of $\epsilon$ with time and the degree of equipartition between relativistic electrons and magnetic field. In the $B_{\rm amp}$ case, if  $\epsilon = 0.002$ instead of $\epsilon = 0.004$, the 1.4 GHz flux density on Day 470 would have been 3.2 mJy rather than 11.5 mJy. Assuming the magnetic field and relativistic electrons are in equipartition, the magnetic field in the shock on Day 470 must have been $<$ 0.009 G to produce less than the observed 1.4 GHz flux density. 

While the predictions for synchrotron luminosity remain uncertain, reasonable parameters imply that significant  synchrotron emission was produced at early times, but it was hidden behind the $\tau = 1$ surface of the ionized Mira wind for most of the evolution of the nova. Despite rich evidence for particle acceleration in V407~Cyg, it is not surprising that we did not detect synchrotron emission in our VLA data. In contrast, if the thermal radio emission had been dominated by the nova ejecta, the synchrotron emission from the forward shock would not have been absorbed, and would have been detectable---yet another reason for dismissing the thermal ejecta scenario investigated in Section \ref{ejecta} as the source of radio emission in V407~Cyg.

\section{Historical Radio Observations of V407~Cyg}\label{hist}

Although only radio non-detections of V407~Cyg have been been reported in the literature for the time period before the 2010 nova outburst (Table \ref{tab:hist}), we used unpublished archival data from the VLA to determine that the source did in fact produce detectable radio emission in 1993, near the beginning of a period of increased activity of the WD \citep{Kolotilov_etal98}. 

The archival VLA data, obtained on 1993 May 14 as part of program AI48 (PI R.~Ivison), were acquired in BnC configuration at X band with 100 MHz bandwidth; they were calibrated using 3C286 and J2007+4029.  With 15 minutes on source, the 1993 data yield a flux density for V407~Cyg of $1.18 \pm 0.07$ mJy at 8.4 GHz. This detection is in contrast to the non-detection obtained four years earlier, when \citet{Seaquist_etal93} found an upper limit on the flux density of $0.06$ mJy at 8.4 GHz. Thus, the strength of the radio emission from V407~Cyg varied by more than a factor of 20 even before the 2010 nova eruption.

The radio luminosity of V407~Cyg probably increased around 1993 because of a flare of ionizing radiation from the WD. From spectroscopic monitoring of V407~Cyg over the last two decades, it is known that the luminosities of optical emission lines vary by factors of $\sim$5 over just a few months. On 1993 May 29, just two weeks after our radio detection, an optical spectrum of V407~Cyg did not show high-excitation signatures of strong symbiotic activity (although it did feature Balmer emission, which had strengthened since 1991; \citealt{Kolotilov_etal98}). By August 1994, emission lines of high ionization states (e.g., [\ion{O}{3}]$\lambda\lambda$4959,5007, \ion{He}{2}$\lambda$4686) were visible in the optical spectrum for the first time since modern observing began of this source in 1984 \citep{Munari_etal94, Kolotilov_etal98}. We hypothesize that this active period of V407~Cyg may have begun in early 1993 and gradually increased in power. The radio appears to precede the strongly symbiotic optical spectrum, but note that in spring 1993, the Mira giant was near the maximum of its pulsation cycle, and the giant's light would have diluted the optical emission line spectrum. In August 1994, the giant is approaching a pulsational minimum, and the relatively faint higher-excitation emission lines would have been more easily visible.

Several years later, in 1998, an outburst was observed in V407~Cyg \citep{Shugarov_etal07}. Unfortunately, no radio light curve exists for the interesting period between 1993 and 1998; V407~Cyg was not observed again with the VLA (except with a shallow NVSS exposure at 1.4 GHz; \citealt{Condon_etal98}, Table \ref{tab:hist}) until the 2010 nova. In addition, no information on the spectral index exists for the 1993 detection of V407~Cyg, so the radio emission mechanism can not be confirmed. In other symbiotic binaries, dramatic increases in radio flux have been associated with collimated mass ejection, which can produce both thermal and synchrotron radio emission \citep[e.g., CH~Cyg, R~Aqr, and Mira;][]{Taylor_etal86, Crocker_etal01, Nichols_etal07, Meaburn_etal09}. While there is, to date, no evidence for fast collimated outflows in optical spectroscopic monitoring, V407~Cyg remains a promising candidate site for the production of synchrotron emission. The dramatic radio variability of V407~Cyg calls for long-term monitoring of the radio flux density and spectral index, in cooperation with synoptic optical spectroscopy, in order to better understand symbiotic activity in this source.

\section{Discussion and Conclusions}

Our VLA light curve, spanning frequencies 1.4--45 GHz and covering 17--770 days after outburst, shows a radio evolution that is unusually luminous and slowly-evolving, compared with other novae. Given the postulated low ejecta mass and lack of detected synchrotron emission, the most plausible source of radio emission from the nova event in V407~Cyg is the ionized wind from the Mira giant. The radio emission from other symbiotic binaries is explained with a similar mechanism, attributing free-free emission to a wind that has been ionized by a hot WD \citep{Seaquist_etal84}. However, the radio emission from V407~Cyg is unique due to the relatively rapid rise and decline of this radio emission. To achieve a sudden rise in the radio light curve requires an abrupt and strong increase in the ionizing flux, and the dramatic fading of the radio light curve requires a mechanism for disrupting or heating the wind on relatively short timescales. We hypothesize that the radio light curve declines as the wind slowly recombines on the outside and is shock-heated from within by the nova blast. The radio light curve provides an estimate of the mass-loss rate of the Mira, $\dot{M}_w \approx 2 \times 10^{-6}~{{\rm M}_{\odot}\ {\rm yr}^{-1}}$ (for $v_w = 10$ km s$^{-1}$ and $d = 3$ kpc). We place a relatively hard lower limit on the mass-loss rate, assuming the wind is completely ionized, of $\dot{M}_w \gtrsim 7 \times 10^{-7}~{{\rm M}_{\odot}\ {\rm yr}^{-1}}$.

\subsection{Different Observational Tracers Imply Discrepant Mira Mass Loss Rates in V407~Cyg}

A few difficulties remain for our claim that $\dot{M}_w \approx 10^{-6}~{{\rm M}_{\odot}\ {\rm yr}^{-1}}$ in V407~Cyg. In a the \emph{Suzaku} X-ray spectrum of V407~Cyg observed on Day 30, non-equilibrium ionization lines are observed, implying an ionization age in the forward shock of $\sim 2 \times 10^{10}$ cm$^{-3}$ s \citep{Nelson_etal12}. This relatively small ionization age translates to an upper limit on the density of circumstellar material at $r_{\rm RG} \approx 10^{15}$ cm of $n_{\rm CSM} \lesssim 10^{5}$ cm$^{-3}$---or a mass-loss rate $\dot{M}_w \lesssim 4 \times 10^{-8}~{{\rm M}_{\odot}\ {\rm yr}^{-1}}$ (an order of magnitude lower than $\dot{M}_w$ determined from the radio; both assume $v_w = 10$ km s$^{-1}$). In addition, studies of the hard X-ray light curve can successfully model the data as shocked Mira wind, but  $\dot{M}_w \approx 10^{-8}-10^{-7}$ M$_{\odot}$ yr$^{-1}$ is required \citep{Orlando_Drake12, Nelson_etal12}. If the wind is too dense, the emission measure of the forward-shocked material will be too high and the ejecta will decelerate too rapidly, making the hard X-ray light curve peak earlier and brighter than observed \citep{Mukai_etal12}. 

A similar effect can be seen in the maximum expansion velocity as a function of time, measured from the wings of H$\alpha$ emission line profiles \citep{Shore_etal11}. If the ratio of the wind density to the ejecta mass is too high, the ejecta will decelerate faster than observed. We compare the redshifted measurements from Shore et al.'s Figure 19 with our simple model of the asymmetric blastwave \citep{Nelson_etal12}. The nova material traveling away from the giant companion maintains the highest velocity, so we compare this direction with the Shore et al.~data. We find that the observed velocity evolution can be reproduced if, for an ejecta mass of M$_{\rm ej} = 10^{-6}$ M$_{\odot}$, the wind mass loss rate is $\dot{M}_w \approx 3 \times 10^{-8}\ {\rm M}_{\odot}\ {\rm yr}^{-1}$ (again, an order of magnitude lower than our radio-determined mass-loss rate). 

On the other hand, the infrared spectral energy distribution of V407~Cyg implies $\dot{M}_w  \approx 8 \times 10^{-7}\ {\rm M}_{\odot}\ {\rm yr}^{-1}$ \citep{Kolotilov_etal98}, in relatively good agreement with our radio estimate. 

All estimates of $\dot{M}_w$ in V407~Cyg assume that the Mira wind is spherically symmetric, which is likely an over-simplification. Simulations show that binary interactions lead to a concentration of circumbinary material in the orbital plane \citep{Walder_etal08, Mohamed_Podsiadlowski11}.
\citet{Orlando_Drake12} find that such a density enhancement is critical for matching the X-ray light curve of V407~Cyg with a small recurrent-nova-like ejecta mass. Radio imaging of V407~Cyg with MERLIN and VLA over the weeks to years following the 2010 nova show structure in the circumbinary medium (\citealt{O'Brien10}, O'Brien et al.~2012, in preparation, Mioduszewski et al.~2012, in preparation). However, the fact that the observed radio luminosity is primarily dependent on the area of $\tau = 1$ surfaces implies that we can not significantly decrease the radio estimate of $\dot{M}_w$ by summoning large-scale structure in the wind. For example, consider the case of a wind funneling most of its mass into a disk-like structure; if the disk is viewed edge-on, the $\tau = 1$ surface will be relatively small and the radio emission will be faint (compared with a spherical wind). The material in the disk will be dense, implying that, if we view it face on, it will only be ionized out to relatively small radius (again producing small $\tau = 1$ surfaces) and will recombine quickly. For these reasons, the most effective way to produce large radio luminosities is to distribute the wind roughly spherically. 

Small-scale ``clumps" in the Mira wind can boost the emission measure of the wind and lead to over-estimates of $\dot{M}_w$ from radio data \citep{Abbott_etal81,Nugis_etal98}. Small-scale clumping, on the other hand, should not affect the IR, X-ray, or H$\alpha$ dynamical estimates of $\dot{M}_w$. SiO masers are detected in the envelope surrounding V407~Cyg \citep{Deguchi_etal05, Deguchi_etal11}, a phenomenon that likely requires an inhomogeneous wind \citep{Gray_etal98}. In addition, the X-ray and radio light curves of the 2010 nova show rapid variations \citep[Figure \ref{lc};][]{Nelson_etal12}, which might be produced by structure in the wind.  This variety of multi-wavelength diagnostics probe different radii in the wind ($\sim 10^{14}$ cm from the Mira for masers; $\sim 10^{15}$ cm for the X-ray light curve; $\sim 10^{15}-10^{16}$ cm for the radio light curve), so we see evidence for inhomogeneities at many radii in the wind of V407~Cyg. Therefore, it is possible that our radio measurement of $\dot{M}_w$ is an over-estimate, and it might be brought into better alignment with the X-ray and H$\alpha$ dynamical estimates by accounting for clumping. However, if the filling factor of the wind is significantly less than unity, the ionization state of the circumbinary material could differ significantly from what we predict in Section \ref{wind}, as it would recombine much more quickly. 

Observed mass loss rates of Mira giants span $\dot{M}_w = {\rm few} \times 10^{-8}-{\rm few} \times 10^{-4}\ {\rm M}_{\odot}\ {\rm yr}^{-1}$, with an average $\langle \dot{M}_w \rangle = 6 \times 10^{-7}$ M$_{\odot} \ {\rm yr}^{-1}$ (\citealt{Knapp_Morris85}; although this sample is not complete for the lowest $\dot{M}_w$ stars). The Mira giant in V407~Cyg has a very long period and is oxygen-rich, so it is more typical of OH/IR stars than Miras \citep{Munari_etal90}. OH/IR sources typically show very dense winds with $\sim10^{-5}\ {\rm M}_{\odot}\ {\rm yr}^{-1}$ \citep[e.g.,][]{Baud_Habing83}, but V407~Cyg is an odd case because dust formation in the wind is suppressed by the hot WD, and therefore the wind-driving mechanism may differ from typical OH/IR stars \citep{Munari_etal90}. Still, we might expect V407~Cyg to occupy the upper end of the Knapp \& Morris range, rather that the lower end---more consistent with the radio and IR mass-loss rates than those implied by the X-ray data and the deceleration of the nova ejecta.

We therefore postulate that the mass-loss rate in V407~Cyg is $\sim10^{-6}\ {\rm M}_{\odot}\ {\rm yr}^{-1}$. One possible solution for reconciling the radio-derived $\dot{M}_w$ with the lower mass-loss rates implied by X-ray non-equilibrium ionization, X-ray emission measure, and shock deceleration is a circumbinary medium that is concentrated in the orbital plane at small radii ($r_{\rm WD} \lesssim 10^{15}$ cm) and becomes more spherically distributed at larger radii ($r_{\rm WD} \gtrsim 10^{15}$ cm). Such a non-spherical wind is in rough keeping with the simulations of circumbinary material by \citet{Walder_etal08}, which show an over-density in the equatorial plane of a factor of $\sim10-100$ for $r_{\rm WD} \lesssim a$, and a flare in the outer circumbinary disk so that the equatorial over-density at larger radius is only a factor of $\sim2-3$. However, this scenario would require significant fine tuning to come into line with the multi-wavelength data on V407~Cyg; the X-ray ionization timescale implies a low density at $\sim 10^{15}$ cm while the 25 GHz light curve measures $\dot{M}_w \approx 10^{-6}\ {\rm M}_{\odot}\ {\rm yr}^{-1}$ at  $\sim 10^{15}$ cm. A sudden radial transition would need to occur in the distribution of circumbinary material, from a ``disky" concentration in the equatorial plane to a near-spherical distribution at a radius of $\sim10^{15}$ cm to reconcile these data. The X-ray and radio data may be brought into better alignment if the binary separation is significantly larger than 17 AU (in which case, the estimated orbital period of 43 years from \citealt{Munari_etal90} would be incorrect). A larger orbital separation would decrease the density of circumbinary material away from the red giant while still allowing for a relatively high $\dot{M}_w$. Our model for the radio light curve would be largely unchanged, but it remains unclear if the X-ray light curve could be explained with such a wide binary separation (\citealt{Orlando_Drake12} kept $a$ fixed at 15.5 AU in their detailed modeling of the X-ray light curve). 

\subsection{Implications for Radio Emission from Novae}
The physical process that we claim is responsible for the radio light curve of V407~Cyg---namely, the time-variable ionization state of the circumbinary environment---has never before been recognized as the primary source of radio emission in a nova outburst. It provides an interesting test case of how dense circumstellar material can affect the evolution of a radio transient. Similar outbursts of thermal radiation, produced by the flash ionization of circumbinary material, might be visible in other novae if significant masses of circumbinary material are prevalent, as suggested by \citet{Williams_Mason10}. Of course, this emission---particularly if it is optically thin---might be superimposed on, or drowned out by, other emission components, like the thermal radiation from the expanding nova ejecta or synchrotron emission from the nova blast. The most similar known nova to V407~Cyg, RS~Oph, was surrounded by a lower density wind; it did not form an absorbing screen following the nova flash, and therefore, synchrotron emission was directly detectable from the nova blast \citep{Hjellming_etal86, Taylor_etal89, O'Brien_etal06, Rupen_etal08}. The recurrent nova T~Pyx is surrounded by an H$\alpha$+[\ion{N}{2}] nebula of mass $\sim \rm{few} \times 10^{-5}$ M$_{\odot}$ \citep{Shara_etal97, Schaefer_etal10} which might have been flash-ionized by the 2010 nova outburst. However, scaling from the H$\alpha$+[\ion{N}{2}] luminosity of this nebula in the months after outburst, we find that the nebula produced a negligible $<$1 $\mu$Jy of thermal radio emission (Nelson et al.~2012, in preparation).  Therefore, V407~Cyg remains the only nova to date that shows a radio light curve dominated by the temperature and ionization structure of its circumbinary medium.

\subsection{Implications for the Progenitors of SNe Ia}
The small nova ejecta mass postulated for V407~Cyg implies that the WD is massive ($\gtrsim$1.25 M$_{\odot}$; \citealt{Mikolajewska10}), making this system an intriguing progenitor candidate for a SN Ia. A symbiotic progenitor like V407~Cyg was conclusively ruled out in the recent case of SN\,2011fe in M101 \citep{Li_etal11, Horesh_etal11, Chomiuk_etal12, Margutti_etal12}, and has also been ruled out for an additional $\sim$15 SNe with deep VLA limits (Chomiuk et al.~2012, in preparation). However, the sustained elusiveness of SN Ia progenitors and subtle diversity of SN Ia explosions has led the community to consider the possibility that SN Ia progenitor systems may be diverse \citep[e.g.,][]{Foley_etal12}. For example, while most SNe Ia show no evidence for hydrogen in their spectra down to very deep limits \citep{Leonard07}, a very small handful of SNe Ia develop luminous H$\alpha$ emission lines in their spectra, indicative of strong interaction with circumstellar material \citep{Hamuy_etal03, Dilday_etal12}. In addition, while some SNe Ia show time-variable blue-shifted \ion{Na}{1} D absorption lines, argued to be evidence of circumstellar material, others do not \citep{Patat_etal07, Patat_etal11b, Sternberg_etal11}. 

Perhaps a fraction of SNe Ia have symbiotic progenitors, with varying delay times between mass loss and SN explosion. For those few WDs that explode into a gas-rich medium like present-day V407~Cyg, the explosion may resemble the H$\alpha$-bright SNe Ia. More commonly, SN Ia-progenitor WDs are surrounded by low-density environments; their immediate environment may have been swept clean by nova blasts that swept up the giant wind \citep{Patat_etal11, Wood-Vasey_Sokoloski06}. However, such a scenario does not describe present day V407~Cyg because, using the classical relation between radio flux density and $\dot{M}_w$ (Equation \ref{eq:mdot}), we derive very similar mass-loss rates for V407~Cyg for frequencies spanning the range 1.4--25 GHz, which are sensitive to the wind density at very different radii ($\sim 10^{15}-10^{16}$ cm). These observations imply that there is no dramatic radial structure in the wind for radii $\gtrsim 10^{15}$ cm, and the Mira envelope follows a smooth $n_{\rm CSM} \propto r^{-2}$ profile. 
%If, as speculated above, the circumbinary material at $< 10^{15}$ cm is concentrated in the orbital plane, the observed evolution of a SN Ia would not be significantly affected, as the SN blast would pass through this zone in $<$1 week. 

Finally, the circumbinary medium around a symbiotic system might have time to expand away and merge with the ambient interstellar medium, if there is a significant delay between accretion and explosion \citep{Justham11, DiStefano_etal11}. V407~Cyg could constitute such a progenitor in $\sim 10^5$ years time. In this case, any blue-shifted \ion{Na}{1} D absorption lines present in the resulting SN Ia spectrum would be the product of the Mira envelope slowly drifting away from its parent binary.

\acknowledgements
We are grateful to N.~Brickhouse, V.~Dwarkadas, D.~Frail, and S.~Shore for helpful discussions, and an anonymous referee for their comments. L.~Chomiuk is a Jansky Fellow of the National Radio Astronomy Observatory. The National Radio Astronomy Observatory is a facility of the National Science Foundation operated under cooperative agreement by Associated Universities, Inc. We are grateful to the EVLA commissioning team, including J.~McMullin, E.~Momjian, L.~Sjouwerman, G.~van Moorsel, and J.~Wrobel, for efficiently scheduling these observations and assisting in their successful completion. We would like to thank ANS Collaboration observers S.~Dallaporta, F.~Castellani, P.~Ochner and L.~Baldinelli for their participation in the photometric monitoring of V407~Cyg.

\bibliography{v407}{}

%\definecolor{Light}{gray}{0.9}
\newpage
\begin{deluxetable}{cccccccccccccc}
\tablewidth{0 pt}
%%\rotate
\tabletypesize{\footnotesize}
\setlength{\tabcolsep}{0.025in}
\tablecaption{ \label{tab:phot}
VLA Observations of V407~Cyg }
\tablehead{Date & $t-t_{0}\tablenotemark{a}$ & Epoch  & Conf. & Freq & $S_\nu$ & Freq & $S_\nu$ & Freq & $S_\nu$ & Freq & $S_\nu$ \\
(UT) & (Days) & & & (GHz) & (mJy) & (GHz) & (mJy) & (GHz) & (mJy) & (GHz) & (mJy)}
\startdata
2010 Mar 27.5 & 17.1 & 1 & D & & & & & 5.0 & $6.39\pm0.35$ & & \\
%\rowcolor{Light} 
2010 Mar 29.8 & 19.0 & 2 & D & 1.5 & $-0.12\pm0.72$ & 1.9 & $1.93\pm0.53$& 5.0 & $7.03\pm0.37$ & & \\

2010 Apr 3.7 & 23.9 & 3 & D & 1.4 & $2.98\pm0.52$ & 1.8 & $3.47\pm0.39$& 4.8 & $6.88\pm0.37$ & 7.0 & $8.81\pm0.45$ \\
 & &  &  & 19.0 & $20.49\pm2.05$ & 25.0 & $24.55\pm2.46$& & & 49.0 & $44.98\pm22.55$ \\

%\rowcolor{Light} 
2010 Apr 8.7 & 28.9 & 4 & D & & & & & 4.8 & $8.99\pm0.47$ & 7.0 & $11.88\pm0.60$ \\
%\rowcolor{Light} 
 & &  &  & 19.0 & $23.65\pm2.37$ & 25.0 & $27.71\pm2.77$ & & & & \\

2010 Apr 12.7 & 32.9 & 5 & D & 1.4 & $4.04\pm0.36$ & 1.8 & $4.76\pm0.38$& 4.8 & $10.68\pm0.55$ & 7.0 & $12.87\pm0.66$ \\
 & &  &  & 19.0 & $29.13\pm2.91$ & 25.0 & $35.28\pm3.53$& 41.0 & $59.89\pm12.01$ & 45.0 & $65.63\pm13.18$ \\

%\rowcolor{Light} 
2010 Apr 15.7 & 35.9 & 6 & D & 1.4 & $3.60\pm0.38$ & 1.8 & $5.47\pm0.41$ & 4.8 & $12.36\pm0.65$ & 7.0 & $14.67\pm0.74$ \\
%\rowcolor{Light} 
 & &  &  & 19.0 & $30.90\pm3.09$ & 25.0 & $37.24\pm3.72$ & & & & \\

2010 Apr 18.7 & 38.9 & 7 & D & 1.4 & $4.78\pm0.38$ & 1.8 & $5.51\pm0.45$ & 4.8 & $12.79\pm0.66$ & 7.0 & $15.59\pm0.79$ \\
  & &  &  & 19.0 & $29.15\pm2.92$ & 25.0 & $34.35\pm3.44$ & & & & \\

%\rowcolor{Light}
 2010 Apr 22.7 & 42.9 & 8 & D & 1.4 & $5.48\pm0.41$ & 1.8 & $6.66\pm0.47$ & 4.8 & $13.69\pm0.71$ & 7.0 & $15.72\pm0.79$ \\
%\rowcolor{Light}
  & &  &  & 19.0 & $28.91\pm2.90$ & 25.0 & $30.32\pm3.05$ & & & & \\

2010 Apr 25.7 & 45.9 & 9 & D & 1.4 & $5.78\pm0.43$ & 1.8 & $6.84\pm0.45$ & 4.8 & $13.96\pm0.70$ & 7.0 & $16.07\pm0.81$ \\
  & &  &  & 19.0 & $30.38\pm3.03$ & 25.0 & $33.86\pm3.39$ & & & & \\

%\rowcolor{Light}
 2010 Apr 29.6 & 49.8 & 10 & D & 1.4 & $5.94\pm0.39$ & 1.8 & $7.38\pm0.47$ & 4.8 & $13.67\pm0.69$ & 7.0 & $16.88\pm0.85$ \\
%\rowcolor{Light}
  & &  &  & 19.0 & $33.03\pm3.30$ & 25.0 & $37.51\pm3.75$ & & & & \\

2010 May 2.6 & 52.8 & 11 & D & 1.4 & $5.76\pm0.40$ & 1.8 & $7.48\pm0.46$ & 4.8 & $13.14\pm0.67$ & 7.0 & $16.10\pm0.81$ \\
  & &  &  & 19.0 & $29.07\pm2.91$ & 25.0 & $34.75\pm3.48$ & & & & \\

%\rowcolor{Light}
 2010 May 7.6 & 57.8 & 12 & D & 1.4 & $6.44\pm0.44$ & 1.8 & $7.66\pm0.51$ & 4.8 & $14.16\pm0.71$ & 7.0 & $17.18\pm0.86$ \\
%\rowcolor{Light}
  & &  &  & 19.0 & $27.77\pm2.78$ & 25.0 & $29.49\pm2.95$ & & & & \\

2010 May 14.5 & 64.7 & 13 & D & 1.4 & $5.36\pm0.39$ & 1.8 & $6.65\pm0.43$& 4.8 & $15.02\pm0.76$ & 7.9 & $19.18\pm0.96$ \\
2010 May 11.6 & 61.8 &  &  & 19.0 & $28.25\pm2.83$ & 25.0 & $30.31\pm3.03$& 41.0 & $51.34\pm10.30$ & 45.0 & $50.66\pm10.16$ \\

%\rowcolor{Light}
 2010 May 16.6 & 66.8 & 14 & D & 1.4 & $5.80\pm0.40$ & 1.8 & $7.57\pm0.46$ & 4.8 & $15.48\pm0.79$ & 7.9 & $19.88\pm1.00$ \\
%\rowcolor{Light}
  & &  &  & 19.0 & $28.75\pm1.44$ & 25.0 & $29.25\pm2.93$ & 41.0 & $44.90\pm9.00$ & 45.0 & $46.31\pm9.28$ \\

2010 May 25.6 & 75.8 & 15 & D & 1.4 & $5.85\pm0.42$ & 1.8 & $8.05\pm0.48$& 4.8 & $17.40\pm0.88$ & 7.9 & $21.71\pm1.09$ \\
 & &  &  & 19.0 & $28.36\pm2.88$ & 25.0 & $28.80\pm2.88$& 41.0 & $39.97\pm8.01$ & 45.0 & $38.40\pm7.71$ \\

%\rowcolor{Light}
 2010 May 29.6 & 79.8 & 16 & D & 1.4 & $5.83\pm0.41$ & 1.8 & $7.53\pm0.47$ & 4.8 & $17.47\pm0.89$ & 7.9 & $20.62\pm1.04$ \\
%\rowcolor{Light}
  & &  &  & 19.0 & $26.72\pm2.67$ & 25.0 & $27.51\pm2.75$ & 41.0 & $37.05\pm7.43$ & 45.0 & $35.88\pm7.20$ \\

2010 Jun 15.5 & 96.7 & 17 & D & 1.4 & $5.61\pm0.38$ & 1.8 & $7.54\pm0.47$& 4.8 & $17.26\pm0.87$ & 7.9 & $20.19\pm1.01$ \\

%\rowcolor{Light}
 2010 Jun 20.6 & 101.8 & 18 & D & 1.4 & $5.76\pm0.44$ & 1.8 & $7.81\pm0.49$& 4.8 & $17.78\pm0.90$ & 7.9 & $19.02\pm0.95$ \\

2010 Jun 21.5 & 102.7 & 19 & D & 1.4 & $5.91\pm0.41$ & 1.8 & $7.38\pm0.45$& 4.8 & $17.45\pm0.88$ & 7.9 & $18.20\pm0.93$ \\
 & &  &  & 19.0 & $23.04\pm2.31$ & 25.0 & $22.47\pm2.25$& 41.0 & $25.94\pm5.23$ & 45.0 & $28.13\pm5.68$ \\
 
%\rowcolor{Light}
 2010 Jul 27.4 & 138.6 & 20 & D & 1.4 & $6.92\pm0.44$ & 1.8 & $8.23\pm0.51$ & 4.8 & $12.93\pm0.65$ & 7.9 & $14.57\pm0.73$ \\
%\rowcolor{Light}
  & &  &  & 19.0 & $14.87\pm1.49$ & 25.0 & $13.63\pm1.37$ & 41.0 & $18.67\pm3.83$ & 45.0 & $18.61\pm3.85$ \\

 2010 Aug 11.4 & 153.6 & 21 & D & 1.4 & $7.92\pm0.52$ & 1.8 & $8.19\pm0.46$& 4.8 & $11.81\pm0.60$ & 7.9 & $12.92\pm0.65$ \\
 & &  &  & 19.0 & $13.26\pm1.33$ & 25.0 & $13.30\pm1.33$& 41.0 & $13.39\pm2.52$ & 45.0 & $13.51\pm2.75$ \\
 
%\rowcolor{Light}
 2010 Aug 26.3 & 168.5 & 22 & D & 1.4 & $5.49\pm0.38$ & 1.8 & $6.39\pm0.40$ & 4.8 & $10.79\pm0.55$ & 7.9 & $11.31\pm0.57$ \\
%\rowcolor{Light}
  & &  &  & 19.0 & $11.87\pm1.19$ & 25.0 & $11.20\pm1.12$ & 41.0 & $14.58\pm2.97$ & 45.0 & $15.12\pm3.08$ \\

 2010 Sep 10.3 & 183.5 & 23 & D & 1.4 & $5.61\pm0.37$ & 1.8 & $6.34\pm0.36$& 4.8 & $9.98\pm0.51$ & 7.9 & $10.16\pm0.51$ \\
 & &  &  & 19.0 & $10.01\pm1.00$ & 25.0 & $9.51\pm0.95$& 41.0 & $9.66\pm2.11$ & 45.0 & $9.09\pm1.98$ \\

%\rowcolor{Light}
 2010 Sep 30.2 & 203.4 & 24 & DnC & 1.4 & $5.58\pm0.31$ & 1.8 & $5.53\pm0.31$ & 4.8 & $9.17\pm0.46$ & 7.9 & $9.10\pm0.46$ \\
%\rowcolor{Light}
  & &  &  & 19.0 & $9.86\pm0.99$ & 25.0 & $10.08\pm1.01$ & 41.0 & $14.19\pm2.91$ & 45.0 & $12.75\pm2.64$ \\

2011 Mar 5.2 & 359.6 & 25 & B & 1.3 & $4.22\pm0.27$ & 1.8 & $4.64\pm0.25$& 5.0 & $5.00\pm0.25$ & 7.4 & $4.81\pm0.24$ \\
 & &  &  & 20.1 & $4.65\pm0.47$ & 25.6 & $4.74\pm0.48$& 41.0 & $5.30\pm1.08$ & 45.0 & $5.45\pm1.11$ \\

%\rowcolor{Light}
 2011 Jun 23.4 & 469.6 & 26 & A & 1.3 & $3.60\pm0.20$ & 1.8 & $4.07\pm0.21$ & 5.0 & $3.27\pm0.17$ & 7.4 & $2.75\pm0.14$ \\
%\rowcolor{Light}
  & &  &  & 20.1 & $2.54\pm0.26$ & 25.6 & $2.62\pm0.27$ & 41.0 & $4.25\pm0.87$ & 45.0 & $4.45\pm0.92$ \\

2011 Nov 4.1 & 603.3 & 27 & D & 1.3 & $3.64\pm0.56$ & 1.8 & $3.86\pm0.35$ & 5.0 & $3.54\pm0.18$ & 7.4 & $3.49\pm0.18$\\
 & &  &  & 20.1 & $3.67\pm0.37$ & 25.6 & $3.83\pm0.39$& 41.0 & $4.32\pm0.87$ & 45.0 & $4.47\pm0.91$ \\

%\rowcolor{Light}
 2012 Jan 08.8 & 669.0 & 28 & DnC & 1.3 & $4.69\pm1.00$ & 1.8 & $3.32\pm0.55$ & 5.0 & $3.21\pm0.17$ & 7.4 & $3.15\pm0.16$ \\
%\rowcolor{Light}
 & &  &  & 20.1 & $3.14\pm0.32$ & 25.6 & $2.99\pm0.30$ & 41.0 & $3.96\pm0.83$ & 45.0 & $5.23\pm1.11$ \\

2012 Apr 18.4 & 769.6 & 29 & C & 1.3 & $2.94\pm0.33$ & 1.8 & $3.35\pm0.29$ & 5.0 & $2.91\pm0.15$ & 7.4 & $2.80\pm0.14$ \\
 & &  &  & 20.1 & $2.91\pm0.29$ & 25.6 & $2.97\pm0.30$ & 41.0 & $2.88\pm0.60$ & 45.0 & $2.66\pm0.59$ \\

\enddata
\tablenotetext{a}{Time since optical discovery of the nova event (2010 March 10.8; \citealt{Munari_etal11}).}
\end{deluxetable}

\newpage
\begin{deluxetable}{cccccccccccc}
\tablewidth{0 pt}
%\rotate
\setlength{\tabcolsep}{0.025in}
\tablecaption{ \label{tab:spind}
Spectral Indices for V407~Cyg }
\tablehead{\colhead{}  & \colhead{} &  \colhead{} &  \multicolumn{2}{c}{1--45 GHz} & \colhead{} & \multicolumn{2}{c}{1--9 GHz} & \colhead{}& \multicolumn{2}{c}{19--45 GHz} \\
\cline{4-5} \cline{7-8} \cline{10-11}\\
Epoch  & $t-t_{0}$ & & $\alpha$ & $\chi^{2}_{\nu}$(dof) & & $\alpha$ & $\chi^{2}_{\nu}$(dof) & & $\alpha$ & $\chi^{2}_{\nu}$(dof)}
\startdata
-1\tablenotemark{b} & 11.7 & & $1.03\pm0.18$ & 2.51(3) & & --- & --- & & $0.98\pm0.26$ & 3.45(2) \\
 2 & 19.0 & & $1.34\pm0.29$ & 0 & & $1.34\pm0.29$  & 0 & & --- & --- \\
 3 & 23.9 & & $0.75\pm0.02$ & 0.28(5) & & $0.68\pm0.01$ & 0.01(2) & & $0.73\pm0.10$ & 0.06(1) \\
 4 & 28.9 & & $0.69\pm0.01$ & 0.06(2) & & $0.74\pm0.19$ & 0 & & $0.58\pm0.52$ & 0 \\
 5 & 32.9 & & $0.76\pm0.02$ & 0.50(6) & & $0.74\pm0.05$ & 0.89(2) &  & $0.94\pm0.08$ & 0.13(2) \\
 6 & 35.9 & & $0.77\pm0.05$ & 2.06(4) & & $0.82\pm0.09$ & 3.27(2) & & $0.68\pm0.52$ & 0 \\
 7 & 38.9 & & $0.70\pm0.03$ & 1.17(4) & & $0.76\pm0.05$ & 0.91(2) & & $0.60\pm0.52$ & 0 \\
 8 & 42.9 & & $0.61\pm0.05$ & 1.42(4) & & $0.66\pm0.05$ & 1.29(2) & & $0.17\pm0.52$ & 0 \\
 9 & 45.9 & & $0.62\pm0.03$ & 0.79(4) & & $0.65\pm0.05$ & 1.14(2) & & $0.39\pm0.51$ & 0 \\
10 & 49.8 & & $0.64\pm0.01$ & 0.17(4) & & $0.64\pm0.02$ & 0.27(2) & & $0.46\pm0.52$ & 0 \\ 
11 & 52.8 & & $0.61\pm0.02$ & 0.40(4) & & $0.61\pm0.04$ & 0.75(2) & & $0.65\pm0.52$ & 0 \\
12 & 57.8 & & $0.55\pm0.03$ & 1.05(4) & & $0.61\pm0.02$ & 0.16(2) & & $0.22\pm0.52$ & 0 \\
13 & 63 & & $0.63\pm0.04$ & 3.01(6) & & $0.74\pm0.05$ & 1.75(2) & & $0.72\pm0.16$ & 0.54(2) \\
14 & 66.8 & & $0.58\pm0.04$ & 3.01(6) & & $0.70\pm0.05$ & 1.47(2) & & $0.56\pm0.17$ & 0.58(2) \\
15 & 75.8 & & $0.56\pm0.06$ & 6.56(6) & & $0.74\pm0.07$ & 3.24(2) & & $0.39\pm0.12$ & 0.30(2) \\
16 & 79.8 & & $0.55\pm0.06$ & 7.05(6) & & $0.73\pm0.08$ & 4.33(2) & & $0.37\pm0.10$ & 0.20(2) \\
17 & 96.7 & & $0.74\pm0.09$ & 5.38(2) & & $0.74\pm0.09$ & 5.38(2) & & --- & --- \\
18 & 101.8 & & $0.68\pm0.12$ & 8.30(2) & & $0.68\pm0.12$ & 8.30(2) & & ---& --- \\
19 & 102.7 & & $0.47\pm0.07$ & 9.26(6) & & $0.67\pm0.11$ & 8.31(2) & & $0.20\pm0.10$ & 0.21(2) \\
20 & 138.6 & & $0.27\pm0.05$ & 4.73(6) & & $0.43\pm0.04$ & 1.17(2) & & $0.27\pm0.21$ & 0.80(2) \\
21 & 153.6 & & $0.19\pm0.04$ & 2.38(6) & & $0.30\pm0.03$ & 0.54(2) & & $-0.03\pm0.05$ & 0.05(2) \\
22 & 168.5 & & $0.28\pm0.05$ & 4.88(6) & & $0.43\pm0.07$ & 2.89(2) & & $0.27\pm0.17$ & 0.54(2) \\
23 & 183.5 & & $0.20\pm0.05$ & 5.41(6) & & $0.35\pm0.06$ & 2.62(2) & & $-0.09\pm0.05$ & 0.04(2) \\
24 & 203.4 & & $0.25\pm0.04$ & 2.82(6) & & $0.33\pm0.07$ & 3.85(2) & & $0.36\pm0.12$ & 0.31(2) \\
25 & 359.6 & & $0.04\pm0.02$ & 0.56(6) & & $0.07\pm0.03$ & 0.82(2) & & $0.19\pm0.03$ & 0.02(2) \\
26 & 469.6 & & $-0.13\pm0.05$ & 3.13(6) & & $-0.17\pm0.07$ & 3.86(2) & & $0.71\pm0.19$ & 0.59(2) \\
27 & 603.3 & & $0.02\pm0.03$ & 0.45(6) & & $-0.05\pm0.02$ & 0.10(2) & & $0.24\pm0.02$ & 0.01(2) \\
28 & 669.0 & & $-0.01\pm0.05$ & 1.15(6) & & $-0.10\pm0.07$ & 0.62(2) & & $0.47\pm0.26$ & 1.04(2) \\
29 & 769.6 & & $-0.03\pm0.02$ & 0.40(6) & & $-0.07\pm0.05$ & 0.61(2) & & $-0.07\pm0.24$ & 0.06(2) \\

\enddata
\tablenotetext{a}{Time since optical V-band maximum (2010 March 10.8; \citealt{Munari_etal11}).}
\tablenotetext{b}{From 2010 March 22/23, as documented in \citet{Nestoras_etal10}.}
\end{deluxetable}

\begin{deluxetable}{ccccc}
\tablewidth{0 pt}
%\rotate
\setlength{\tabcolsep}{0.025in}
\tablecaption{ \label{tab:pach}
Constants for Calculating Synchrotron Emission \citep{Pacholczyk70}}
\tablehead{$p$ & & $c_5$ & & $c_6$}
\startdata
2.0 &  & $7.52 \times 10^{-24}$ &   & $7.97 \times 10^{-41}$\\
2.5 &  & $9.68 \times 10^{-24}$ &   & $8.10 \times 10^{-41}$\\
3.0 &  & $1.37 \times 10^{-23}$ &   & $8.61 \times 10^{-41}$\\
\enddata
\end{deluxetable} 

\begin{deluxetable}{ccccc}
\tablewidth{0 pt}
%\rotate
\setlength{\tabcolsep}{0.025in}
\tablecaption{ \label{tab:hist}
Historical Radio Observations of V407~Cyg}
\tablehead{Date & Frequency & $S_\nu$ & Facility & Reference\\
(UT) & (GHz) & (mJy) & &}
\startdata
1982 Jul 13 & 4.9 & $< 0.96$ & VLA & \citet{Seaquist_etal84}\\
1988 Oct 16 & 5.0 & $< 1.72$ & BBI\tablenotemark{a} & \citet{Ivison_etal91}\\
1989 Dec 22 & 8.4 & $< 0.06$ & VLA & \citet{Seaquist_etal93}\\
1992 Oct 10 & 270 & $< 114$ & JCMT & \citet{Ivison_etal95}\\
1993 May 14 & 8.4 & $1.18\pm0.07$ & VLA & This Work \\
1996 Aug 18 & 1.4 & $< 2.77$& VLA & \citet{Condon_etal98} \\ 
\enddata
\tablenotetext{a}{The Jodrell Bank Lowell-Mark 2 Broad Band Interferometer.}
\end{deluxetable} 

\end{document}